\documentclass[a4paper,11pt]{article}
\pdfoutput=1 

\usepackage{jheppub} 

\usepackage{amsmath, amsfonts, amsthm, amssymb, graphicx, color, hyperref, slashed, braket}
\usepackage{subfigure}
\usepackage{pstool}
\usepackage{graphicx}
\usepackage{float} 
\usepackage[utf8]{inputenc}
\usepackage[normalem]{ulem}
\usepackage[english]{babel}
\usepackage{blkarray}

\allowdisplaybreaks[3]

\def \bal#1\eal  {\begin{align} #1 \end{align}}
\def\({\left(}
\def\){\right)}
\def\[{\left[}
\def\]{\right]}
\def\<{\left\langle}
\def\>{\right\rangle}
\def\d{\mathrm{d}}

\newcommand{\td}{\tilde}
\newcommand{\eref}[1]{Eq.~(\ref{#1})}
\newcommand{\f}[2]{\frac{#1}{#2}}
\newcommand{\bim} {\begin{itemize}[noitemsep]}   
   
\newcommand{\eim}{\end{itemize}}
\newcommand{\be} {\begin{equation}} 
\newcommand{\ee} {\end{equation}}
\newcommand{\bc}{\begin{center}}   
\newcommand{\ec}{\end{center}}

\newcommand{\nn} {\nonumber\\}

\newcommand{\nb}{\normalsize\tt}

\newcommand{\marrow}{~~\Longrightarrow~~}

\newcommand{\ie}{{\it i.e.,}~}

\newcommand{\im}{{\rm Im}}

\newcommand{\mc} {\mathcal}

\newcommand{\ai}{{\alpha}}

\newcommand{\li}{{\lambda}}

\newcommand{\epi}{\epsilon}
\newcommand{\thi}{\theta}
\newcommand{\Gi}{\Gamma}
\newcommand{\Li}{\Lambda}
\title{Triple crossing positivity bounds for multi-field\! theories}

\author[a]{Zong-Zhe Du,}
\author[b,c]{Cen Zhang}
\author[a,d]{and Shuang-Yong Zhou}

\affiliation[a]{Interdisciplinary Center for Theoretical Study, University of Science and Technology of China, Hefei, Anhui 230026, China}
\affiliation[b]{Institute for High Energy Physics, and School of Physical Sciences, University
of Chinese Academy of Sciences, Beijing 100049, China}
\affiliation[c]{Center for High Energy Physics, Peking University, Beijing 100871, China}
\affiliation[d]{Peng Huanwu Center for Fundamental Theory, Hefei, Anhui 230026, China}

\emailAdd{mmmao@mail.ustc.edu.cn}
\emailAdd{zhoushy@ustc.edu.cn}

\preprint{\footnotesize USTC-ICTS/PCFT-21-42}

\date{\today}

\abstract{  
We develop a formalism to extract triple crossing symmetric positivity bounds for effective field theories with multiple degrees of freedom, by making use of $su$ symmetric dispersion relations supplemented with positivity of the partial waves, $st$ null constraints and the generalized optical theorem. This generalizes the convex cone approach to constrain the $s^2$ coefficient space to higher orders. Optimal positive bounds can be extracted by semi-definite programs with a continuous decision variable, compared with linear programs for the case of a single field. As an example, we explicitly compute the positivity constraints on bi-scalar theories, and find all the Wilson coefficients can be constrained in a finite region, including the coefficients with odd powers of $s$, which are absent in the single scalar case.
}

\begin{document}
\maketitle
\flushbottom

\section{Introduction and summary}

Recently, there has been significant progress in understanding the consistent parameter spaces of effective field theories (EFTs) in the form of positivity bounds. These are constraints on the Wilson coefficients of an EFT, or more generally on some physical observables derived from the EFT scattering amplitudes, and arise from simply assuming the EFT's UV completion is consistent with fundamental principles of S-matrix such as causality/analyticity and unitarity. The positivity bounds can be quite stringent, often eliminating large chunks of the naive parameter space, which highlights the fact that not everything consistent with the symmetries can be a valid EFT.
 
Applying the optical theorem to the dispersion relation for an identical scalar amplitude, one can derive the forward positivity bound which states that the Wilson coefficient in front of $s^2$ should be positive \cite{Adams:2006sv} (see also earlier works \cite{Pham:1985cr, Ananthanarayan:1994hf}), $s,t,u$ being the standard Mandelstam variables. There are a few directions to generalize the scope and strength of positivity bounds for 2-to-2 scattering. First of all, the same argument for the $s^2$ bound can also be used to infer that the coefficients in front of all even powers of $s$ are all positive, if the theory is weakly coupled at least in the IR. But there is more to the forward dispersion relation than meets the eye of the optical theorem's positivity. In fact, the Hankel matrix of all the coefficients of even $s$ powers are also positive definite \cite{Arkani-Hamed:2020blm}, as can be seen from the connection to the (mathematical) moment problems \cite{Arkani-Hamed:2020blm, Bellazzini:2020cot}. Once the $t$ expansion of the dispersion relation is also included, the structure of positivity bounds becomes much richer. By using the positivity of derivatives of the amplitude's absorptive part and relaxing the UV scale of the dispersive integrand, Ref \cite{deRham:2017avq} derived an infinite tower of positivity bounds away from the forward limit that can be cast as a recurrence relation involving $s$ and $t$ derivatives (see \cite{Manohar:2008tc, Nicolis:2009qm, Bellazzini:2016xrt} for earlier works on going beyond the forward limit). These recurrence bounds have also been generalized to massive particles with spin using the transversity basis for the external polarizations in which $su$ crossing relations are (semi-)diagonalized \cite{deRham:2017zjm}.

However, these recurrence positivity bounds only utilized the $su$ crossing symmetry that is inherent in the twice subtracted dispersion relation. It has been realized that huge gains can be profited if one imposes $st$ crossing symmetry on the $su$ symmetric dispersion relation, which allows us to bound all the coefficients in the $s,t$ expansion both from above and below \cite{Tolley:2020gtv, Caron-Huot:2020cmc}. Specifically, the $st$ crossing symmetry implies that certain linear combinations of the Wilson coefficients in the $su$ symmetric expansion of the EFT amplitude have to vanish, which, by substituting in the sum rules from the $su$ symmetric dispersion relation, gives rise to an infinite series of null constraints on the dispersive integrals. These null constraints can be added into the sum rules of the Wilson coefficients to form linear programs to extract the optimal bounds from the positivity of the UV spectral function. These linear programs involve a continuous decision variable, which nevertheless can be efficiently solved by the publicly available {\nb SDPB} package \cite{Simmons-Duffin:2015qma}. For the $s^2$ coefficient, Ref \cite{Caron-Huot:2020cmc} was also able to derive an upper bound in terms of the cutoff, thanks to the upper bound of partial wave unitarity and the first null constraint. For the case of a single scalar field, these new triple crossing symmetric bounds can restrict the dimensionless Wilson coefficients to be parametrically order $\mc{O}(1)$. This is of course something one naturally expects, as can be seen in many examples by explicitly integrating out the heavy fields from the UV models, but these new bounds put the naturalness argument on a rigorous and concrete footing, barring the possibility of a potential accidental large coupling. 

An alternative way to use the triple permutation symmetry of a scalar amplitude is to start with a dispersion relation that is triple symmetric \cite{Sinha:2020win} (based on an earlier work \cite{Auberson:1972prg}), and then locality enforces an alternative set of null constraints on the triple symmetric dispersion relation. Yet, another way to extract the same positivity bounds \cite{Chiang:2021ziz} is to first perform a general linear rotation to simplify the partial wave expansion in the dispersion relation and then convert the problem to a bi-variate moment problem that is well studied in mathematics. Whether a point in the parameter space satisfies the positivity bounds can then be determined by checking positive definiteness of a series of coefficient matrices, and the null constraints in this way can be imposed at the level of Wilson coefficients, that is, by slicing out the triple crossing symmetric subspace of the allowed bounds.

An interesting application of these triple crossing symmetric bounds is that, while the forward positivity bound can marginally rule out \cite{Adams:2006sv} massless Galileon theory \cite{Nicolis:2008in}, which arises as decoupling limits of several gravitational models (see {\it e.g.,} \cite{deRham:2016nuf}), the new bounds can now effectively rule out Galileon theories where the Galileon symmetry is weakly broken, as these bounds dictate that the weakly broken scale must be parametrically close to the cutoff scale in these theories \cite{Tolley:2020gtv}. On the other hand, full-blown applications of positivity bounds in gravitational theories requires a judicial treatment of the $t$ channel pole, which survives the twice subtraction and whose singularity in the forward limit is balanced out by the divergence in the dispersive integral. By transforming to the impact parameter space, it has been shown that the forward singularity can be overcome, carving out some sharp boundaries for the swampland \cite{Caron-Huot:2021rmr}. See  \cite{Bellazzini:2015cra, Cheung:2016yqr, Bonifacio:2016wcb, deRham:2017imi, Bellazzini:2017fep, deRham:2018qqo, Bonifacio:2018vzv, Melville:2019wyy, deRham:2019ctd, Alberte:2019xfh, Alberte:2020bdz, Chen:2019qvr, Huang:2020nqy, Alberte:2020jsk, Tokuda:2020mlf, Wang:2020xlt, Herrero-Valea:2019hde, Herrero-Valea:2020wxz, deRham:2021fpu, Traykova:2021hbr, Bern:2021ppb, Arkani-Hamed:2021ajd, Davis:2021oce} for some other interesting discussions of positivity bounds in gravity and cosmology.

While the above positivity bounds for a single scalar field presents a significant step towards a better understanding of the structure of the parameter space of EFTs, our universe is more complex than just a single scalar field. We typically encounter EFTs with many degrees of freedom.  In the absence of any new particle signals at the LHC, the EFT approach to parameterize possible Beyond the Standard Model physics has become increasingly popular. In the Standard Model EFT (SMEFT), for example, there are a large number of low energy modes. How to optimally handle many degrees of freedom adds a whole new dimension to the problem of extracting the positivity bounds in a generic EFT. Recently, progress has been made in this direction for the lowest order $s^2$ coefficients, the positivity bounds on which are of course important phenomenologically. 

If there are many modes in a low energy EFT, a simple generalization of the elastic positivity bounds is to linearly superpose the different modes to get elastic positivity bounds for the superposed states. However, this does not produce the strongest positivity bounds, as it misses positivity bounds coming from considering elastic amplitudes between ``entangled states'' \cite{Zhang:2020jyn, Li:2021lpe}. By using generalized optical theorem, one can see that the $s^2$ coefficients of the multi-field amplitudes form a convex cone, whose extremal rays correspond to irreps of the EFT symmetries or one particle UV states projected down to the EFT symmetries \cite{Zhang:2020jyn}. This highlights the importance of positivity bounds in inverse-engineering the UV completion of SMEFT and the importance of higher order operators in Beyond the Standard Model phenomenology. Furthermore, the dual of the amplitude cone is a spectrahedron, and thus finding the strongest positivity bounds can be turned into a (normal) semi-definite program (SDP) \cite{Li:2021lpe}, which can be efficiently solved by many widely available algorithms. Other applications of positivity bounds in SMEFT can be found in  \cite{Zhang:2018shp, Bi:2019phv, Yamashita:2020gtt, Fuks:2020ujk, Zhang:2020jyn, Gu:2020ldn, Li:2021lpe, Vecchi:2007na, Bellazzini:2018paj, Remmen:2019cyz, Trott:2020ebl, Remmen:2020vts, Bonnefoy:2020yee, Chala:2021wpj}. Also, see \cite{Distler:2006if, Remmen:2020uze, Grall:2021xxm, Davighi:2021osh, Haldar:2021rri, Raman:2021pkf, Gopakumar:2021dvg, Zahed:2021ffy, Kundu:2021qpi} for a few other generalizations of positivity bounds, and \cite{Paulos:2016fap, Guerrieri:2020bto, Hebbar:2020ukp, Guerrieri:2021tak} and reference therein for recent progress in S-matrix bootstrap, which overlaps the development of positivity bounds.
 
In this paper, we initiate the study of positivity bounds on higher order Wilson coefficients for theories with multiple degrees of freedom, incorporating both the triple crossing and convex cone approach, to better the understanding of the parameter space structure of multi-field EFTs. We will focus on theories with scalar fields only, which have simpler partial wave expansions. For the multi-field case, the dispersion relation contains a spectral tensor that is indexed by the four external particles and is not necessarily positive for generic combinations of the particles. To extract the optimal bounds, we take a convex geometry approach to view this spectral tensor as living in a convex cone. A key observation is that the dual cone of the spectral tensor cone is a spectrahedron. This allows us to formulate an SDP with only one continuous decision variable to extract the optimal positivity bounds, which can again be efficiently solved by the {\nb SDPB} package. This is compared to the single scalar case which can be formulated as linear programs with one continuous variable. For both cases, the continuous variable comes from the dispersive integral and is associated with possible scales of the UV states. On the other hand, this is also similar to the $s^2$ convex cone, but now the spectrahedron depends on the partial waves, the UV scale and the orders of the $v=s+t/2$ and $t$ expansion. 

Due to the multi-field structure, the generic $st$ null constraints are now more complex, as they are now indexed by the external particles and the external particle indices are swapped for some terms in the constraints. Nevertheless, a convenient way to add them to the SDP is essentially to invoke something similar to the dual cone for the Wilson coefficients. As the null constraints are equalities, instead of inequalities that are used to define the dual cone in convex geometry, the structure that is introduced is really just the boundary of the dual cone. Another way to use our formalism, which is not explicitly demonstrated in the paper, is to perform the SDPs without adding the dispersive null constraints. Instead, the null constraints are to be imposed as equalities on the Wilson coefficients directly. That is, one restricts the outcomes of these SDPs to the subspace defined by the null constraints in the coefficient space. However, this typically involves computing SDPs with quite a few coefficients, as even the lowest null constraints in a multi-field theory already contain quite a few coefficients. If we are faced with a problem that involves only a couple of particular coefficients, the approach explicitly implemented in this paper is much more efficient.

As an illustration, we apply the formalism to constrain the Wilson coefficients of bi-scalar theories endowed with some discrete symmetries. We explore the geometric shapes for the $v^2$ and $v^2t$ coefficients and also compute two-sided bounds for higher order coefficients, which are obtained agnostic about the other coefficients. To obtain these two-sided bounds, it is essential to take into account the finiteness of the $v^2$ coefficients. Different from the single scalar case, we now also have coefficients with odd powers of $s$, and we find that these coefficients are also bounded from both sides, despite that its associated (raw) spectral tensor does not form a salient cone. All in all, the triple crossing bounds can be used to constrain the Wilson coefficients of a multi-field EFT in a finite region near the origin.

The paper is organized as follows. In Section \ref{sec:sumrules}, we derive the $su$ symmetric dispersion relation and expand the relation to get sum rules for all the Wilson coefficients, and then we impose $st$ crossing symmetry to get the null constraints on the dispersion relation. In Section \ref{sec:sdp}, we formulate the extraction of triple crossing symmetric positivity bounds as SDPs, show how these programs can be implemented with {\nb SDPB}. In Section \ref{sec:bi-scalar}, for a simple example, we explicitly calculate the triple crossing positivity bounds for bi-scalar theory with the double $\mathbb{Z}_2$ symmetry and the $\mathbb{Z}_2$ symmetry, with the numerical results presented in five figures and one table. In Appendix \ref{sec:EF}, we list the explicit expressions for a few quantities used in the first few null constraints for a quick reference. A briefly discussion on generalization to the case with massive fields is presented in Appendix \ref{app:massive}.

\section{Sum rules for multi-fields}
\label{sec:sumrules}

In many circumstances, EFTs contain multiple low energy modes in their spectrum so as to reproduce our phenomenal world. Let us suppose that there are $N$ light modes in a $D$ dimensional low energy EFT. Consider an EFT with a large hierarchy between the cutoff and the masses of the modes, $\Lambda\gg m_i$, so that it is a good approximation to take the massless limit $m_i\to 0$, and therefore we have the following simple kinematic relations
\be
s+t+u= 0,~~~\cos\thi = 1+\f{2t}{s} ,
\ee
where $s = -(p_1+p_2)^2,t=-(p_1-p_3)^2,u=-(p_1-p_4)^2$ with $p_1,p_2,p_3,p_4$ being the external momenta, and $\thi$ is the scattering angle between particle 1 and 3. For simplicity, we will consider multi-scalar theories as our examples, as their partial wave expansion is easier to perform explicitly. For massive scalars, whose crossing relations are still trivial, a generalization to the case with the same mass is also straightforward --- see Appendix \ref{app:massive}.  We will assume that the EFT is weakly coupled and loop contributions are suppressed below the cutoff, so we can focus on the leading tree level amplitudes. For tree level amplitudes, the positivity bounds can be directly expressed in terms of the Wilson coefficients, while for the loop amplitudes it might be more convenient to express the bounds in terms of observables derived from the amplitude.

\subsection{$su$ symmetric dispersion relation}

To derive positivity bounds for an EFT, we make use of the $su$ symmetric dispersion relation, whose existence reflects unitarity, analyticity, locality and crossing symmetry of the underlying UV amplitude. Consider UV scattering amplitude $A_{ijkl}(s, t)$ for process $ij\to kl$, where $i,j,k,l=1,2,...,N$ label different low energy modes. Following the same steps as the case of a single scalar (see, eg, \cite{deRham:2017avq}), particularly utilizing Cauchy's integral formula for $A_{ijkl}(s, t)$ in the complex $s$ plane for fixed $t$ along with the Froissart-Martin bound \cite{Froissart:1961ux, Martin:1962rt} and $su$ crossing symmetry, we can express the amplitude as a dispersive integral of the absorptive part of the amplitude over $\mu>\Lambda^2$:
\bal
 B_{ijkl}(s,t) &\equiv A_{ijkl}(s,t)- \frac{\li_{ijkl}}{-s}  - \frac{\li_{ijkl}}{ -t} - \frac{\li_{ijkl}}{ -u}
\\
&=  a^{(0)}_{ijkl}(t)+a^{(1)}_{ijkl}(t)s+ \int^\infty_{\Li^2}  \frac{\d \mu}{\pi (\mu-\mu_p)^2} \[ \frac{(s-\mu_p)^2}{\mu-s}{\rm Abs}\, A_{ijkl}(\mu,t)+\frac{(u-\mu_p)^2}{\mu-u} {\rm Abs}\, A_{ilkj}(\mu,t)  \]  \nonumber
\\
&=  a^{(0)}_{ijkl}(t)+a^{(1)}_{ijkl}(t)s+ \int^\infty_{\Li^2}  \frac{\d \mu}{\pi (\mu+\f{t}2)^2} \[ \frac{(s+\f{t}2)^2}{\mu-s}{\rm Abs}\, A_{ijkl}(\mu,t)+\frac{(u+\f{t}2)^2}{\mu-u} {\rm Abs}\, A_{ilkj}(\mu,t)  \]  , \nonumber
\eal
where $\Lambda$ is the scale at which the lowest heavy modes come in, identified with the cutoff here for simplicity and the absorptive part is defined as ${\rm Abs}\,A(\mu,t)=\f1{2i}{\rm Disc}\,A(\mu,t)=\f1{2i}\big[A(\mu+i\epi,t)-A(\mu-i\epi,t)\big]$ with $\epi\to 0^+$. When $A_{ijkl}$ is time reversal invariant, the absorptive part reduces to the imaginary part ${\rm Abs}\, A_{ijkl}(\mu,t)=\im A_{ijkl}(\mu,t)$. This is the so-called twice subtracted dispersion relation with the subtraction terms $a^{(0)}_{ijkl}(t)$ and $a^{(1)}_{ijkl}(t)$, which are needed because the Froissart-Martin bound, derived from unitarity, analyticity and locality/polynomial boundedness, can only constrain the UV behavior of ${\rm Abs}\, A_{ilkj}(s,t)$ to diverge slower than $s^2$. In the last line of the equation above, we have chosen the subtraction point at $\mu_p=-{t}/2$ so that $(s+\f{t}2)^2=(u+\f{t}2)^2$ can factor out of the whole dispersive integral, which allows us to define a convenient SDP later. Since $B_{ijkl}(s,t)$ is $su$ symmetric, we have $B_{ijkl}(s,t) =  B_{ilkj}(u,t)$, so $ a^{(0)}_{ijkl}(t)$ and $a^{(1)}_{ijkl}(t)$ must satisfy the following relations 
\be
a^{(1)}_{ijkl}(t) = -a^{(1)}_{ilkj}(t), ~~~~ a^{(0)}_{ijkl}(t)= a^{(0)}_{ilkj}(t) -a^{(1)}_{ilkj}(t) t .
\ee
We have chosen the scalar fields to be in a self-conjugate basis for simplicity, but the formalism below can be easily generalized for a general basis, in which one should additionally conjugate the particle after crossing. For example, the crossing symmetry would be $B_{ijkl}(s,t) =  B_{i\bar{l}k\bar{j}}(u,t)$ and the $u$ channel part of the dispersion relation would have ${\rm Abs}\, A_{i\bar{l}k\bar{j}}(\mu,t)$, where $\bar{j}$ and $\bar{l}$ stands for the conjugate particle of $j$ and $l$ respectively.  

For our later convenience, it is useful to introduce a new momentum invariant $v$ to replace the $s$ variable
\be
v=s+\f{t}2   ,
\ee
and then the $s\leftrightarrow u$ crossing symmetry simply becomes $v\leftrightarrow -v$:
\be
B_{ijkl}(s,t) =  B_{ilkj}(u,t) \marrow   \tilde B_{ijkl}\(v,t \) =  \tilde B_{ilkj}\(-v,t\)  ,
\ee
where we have defined $\tilde B_{ijkl}(v,t) \equiv B_{ijkl}(s,t)$. Then we can write the dispersion relation as
\bal
\tilde B_{ijkl}(v,t)  = \tilde a^{(0)}_{ijkl}(t)+a^{(1)}_{ijkl}(t)v+ v^2 \int^\infty_{\Li^2}  \frac{\d \mu}{\pi (\mu+\f{t}2)^2} \[ \frac{{\rm Abs}\, A_{ijkl}(\mu,t)}{\mu+\f{t}2-v}+\frac{{\rm Abs}\, A_{ilkj}(\mu,t)}{\mu+\f{t}2+v}   \]       .
\eal
For elastic scattering, \ie when $i=k,~j=l$, we can also factor out ${\rm Abs}\, A_{ijij}(\mu,t)$ in the integrand and have $a^{(1)}_{ijij}(t) = -a^{(1)}_{ijij}(t)=0$, which means that there will be only even powers of $v$ on the both sides of the dispersion relation, as is the case for scattering between identical particles. For inelastic scattering amplitudes, odd powers of $v$ are generally present, and we will see that the Wilson coefficients of these terms also have two-sided bounds, despite their individual spectral tensors not forming salient cones. The remarkable feature of the dispersion relation is that on the left hand side the quantity $B_{ijkl}(s,t)$ can be well approximated by the EFT computations for $s,t\ll \Lambda^2$, while the right hand side relies on the absorptive part of the amplitude from the high energy UV theory, noting that the integration goes all the way up to infinity. In other words, the dispersion relation is a tool for us to use some salient properties of UV physics to constrain the EFT in the IR.

The absorptive part of the amplitude can be expanded by partial waves
\be
{\rm Abs} \, A_{ijkl}(\mu,t)=  \frac{2^{4\ai+2}\pi^\ai \Gamma(\ai)}{\mu^{\ai-\f12}}  \sum_{\ell=0}^\infty (2\ell+2\ai)C^{(\ai)}_{\ell}\(1+\f{2t}{\mu}\) {\rm Abs}\, a^{ijkl}_\ell(\mu) , ~~~\ai = \frac{D-3}2   ,
\ee
where the Gamma function $\Gamma(\ai)$ is positive for $D\geq 3$ and $C^{(\ai)}_{\ell}$ is the Gegenbauer polynomial, the $D$ dimensional generalization of the Legendre polynomial in 4D. In a scattering process, angular momenta being conserved implies that unitarity can be applied to individual partial waves. A direct consequence of partial wave unitarity is the generalized optical theorem, which for partial wave $\ell$ means
\be
\label{absaijkl}
{\rm Abs}\, a^{ijkl}_\ell = \sum_X a^{ij\to X}_{\ell} (a^{kl\to X}_{\ell})^*    ,
\ee
where $X$ denotes all possible intermediate states. For later convenience, we can absorb the positive factor in the expansion into the $ij\to X$ partial wave amplitude and define
\bal
\label{mdef0}
 m_\ell^{ij}(\mu)  &\equiv  \[ \frac{2^{4\ai+2}\pi^{\ai-1} \Gamma(\ai)}{ \mu^{\ai-\f12}}   (2\ell+2\ai)C_\ell^{(\ai)}(1) \]^{1/2} a^{ij\to X}_{\ell}(\mu)   .
\eal
(We have included the constant $C_\ell^{(\ai)}(1)= {\Gamma(\ell+D-3)}/({\Gamma(D-3) \Gamma(\ell+1)})>0$ for $D\geq 4$, as we shall soon also expand $C^{(\ai)}_{\ell}(1+{2t}/{\mu})$ in terms of $t$.) As we want to extract positivity bounds that are independent of the UV models, we shall take the amplitude from $ij$ to $X$, $m_\ell^{ij}(\mu)$, to be arbitrary for every $\mu$ and $\ell$, except for certain symmetries between the indices $i$ and $j$ that are already known in the low energy EFT --- the exact values of $m_\ell^{ij}(\mu)$ are ultimately determined by specifics of the UV completion. Defining a short hand notation 
\be
\<\; \cdots\phantom{\Big|}\!\!\> \equiv \sum_{X,\ell} \int^\infty_{\Li^2} \d \mu\; \Big(\cdots\Big)   ,
\ee
the dispersion relation can be written as
\bal
\label{usefulDispersion}
&~~~~~\tilde B_{ijkl}(v,t) -\tilde a^{(0)}_{ijkl}(t)-a^{(1)}_{ijkl}(t)v
\\
&= \<  \[ \frac{v^2 m_\ell^{ij} (m_\ell^{kl})^*}{(\mu+\f{t}2)^2(\mu+\f{t}2-v)}  +  \frac{v^2 m_\ell^{il} (m_\ell^{kj})^*}{(\mu+\f{t}2)^2(\mu+\f{t}2+v)}  \]  \f1{C_\ell^{(\ai)}(1)} C^{(\ai)}_{\ell}\(1+\f{2t}{\mu}\) \>   .
\eal
We have subtracted the $v^0$ and $v^1$ terms (\ie the $\td a^{(0)}_{ijkl}(t)$ and $a^{(1)}_{ijkl}(t)$ terms) from $\tilde B_{ijkl}(s,t)$, so the expansion on the right hand side starts from $v^2$. As we shall see shortly, $\td a^{(0)}_{ijkl}(t)$ and $a^{(1)}_{ijkl}(t)$ can also be expressed in terms of dispersive integrals by imposing the $st$ crossing symmetry $B_{ijkl}(s,t)=B_{ikjl}(t,s)$ on the $su$ symmetric dispersion relation. Expanding both sides of \eref{usefulDispersion} on $v$ and $t$ and re-ordering the summations appropriately, we can get
\bal
\label{expandDisRel}
\sum^\infty_{m=2}\sum^\infty_{n=0} c_{ijkl}^{m,n} v^m t^n&=  \tilde B_{ijkl}(v,t) - \tilde a^{(0)}_{ijkl}(t)-a^{(1)}_{ijkl}(t)v
\\
 &= \sum_{m=2}^{\infty} \sum_{n=0}^\infty \<    \[m_\ell^{ij} (m_\ell^{kl})^* +(-1)^m  m_\ell^{il} (m_\ell^{kj})^*     \]\sum_{p=0}^n  \f{L^{p}_\ell H^{n-p}_{m+1}}{\mu^{m+n+1}} \>  v^m t^n   ,
\eal
where we have defined the Taylor expansion notation for the Gegenbauer polynomials
\bal
L^n_\ell &\equiv\f{2^n}{n!C^{(\ai)}_{\ell}(1)}\left. \frac{\d^n}{\d x^n}C^{(\ai)}_{\ell}(x)\right|_{x=1} 
=\frac{ \Gamma (\ell+n+2 \ai )\Gamma (\ell+1) \Gamma \left(\ai +\frac{1}{2}\right)}{ n! \Gamma (\ell+2 \ai )\Gamma (\ell-n+1) \Gamma \left(n+\ai +\frac{1}{2}\right)}   
\geq 0    ,
\eal
and the Taylor expansion notation for $(\mu+\f{t}2)^{-m-1}$,
\be
H_{m+1}^q\equiv \f{\Gamma(m+q+1)}{(-2)^q\Gamma(q+1)\Gamma(m+1)}   .
\ee
Matching the coefficients of powers of $v$ and $t$ on the two sides of \eref{expandDisRel} gives a set of sum rules for the Wilson coefficients
\be
\label{csumrules}
c_{ijkl}^{m,n} = \< C_{ijkl}^{m,n}\> \equiv \<   \[m_\ell^{ij} (m_\ell^{kl})^* +(-1)^m  m_\ell^{il} (m_\ell^{kj})^*     \] \f{C_\ell^{m,n}}{\mu^{m+n+1}}  \>,~~m\geq 2,n\geq 0  ,
\ee
where we have further defined
\be
C_\ell^{m,n} \equiv \sum_{p=0}^n L^{p}_\ell H^{n-p}_{m+1}   .
\ee
The first few $C_\ell^{m,n}$ in 4D are explicitly given by
\be
C_\ell^{2,0}= 1,~~C_\ell^{2,1}=\ell^2+\ell-\f32,~~ C_\ell^{2,2}=\f14 (\ell^4+2 \ell^3-7 \ell^2-8 \ell+6),~~C_\ell^{3,1}=\ell^2+\ell-2   .
\ee 
From these sum rules, we can see that even $m=2h$ and odd $m=2h+1$ Wilson coefficients $c_{ijkl}^{m,n}$ satisfy different symmetries when exchanging $j\leftrightarrow l$
\be
\label{cijklsymm1}
c_{ijkl}^{2h,n} = c_{ilkj}^{2h,n},~~~c_{ijkl}^{2h+1,n} = -c_{ilkj}^{2h+1,n}   .
\ee 

If the theory is time reversal invariant, then we can restrict to the case where $m^{ij}_\ell$ are real numbers. To see this, let us split $m^{ij}_\ell$ into $m^{ij}_{\ell,R}+i \,m^{ij}_{\ell,I}$, and we have
\be
\label{mmRealNot}
m^{ij}_\ell (m^{kl}_\ell)^*+(j\leftrightarrow l) =  m^{ij}_{\ell,R} m^{kl}_{\ell,R}+ m^{ij}_{\ell,I} m^{kl}_{\ell,I} + i \big( m^{ij}_{\ell,I} m^{kl}_{\ell,R}- m^{ij}_{\ell,R} m^{kl}_{\ell,I} \big) + (j\leftrightarrow l)   .
\ee
If the amplitude is invariant under time reversal, then we have $m^{ij}_\ell (m^{kl}_\ell)^*+(j\leftrightarrow l)=m^{kl}_\ell (m^{ij}_\ell)^*+(j\leftrightarrow l)$, which means that the imaginary part  of \eref{mmRealNot} vanishes. For this case, if we re-define 
\be
\<\; \cdots\phantom{\Big|}\!\!\> \equiv \sum_{X',\ell} \int^\infty_{\Li^2} \d \mu\; \Big(\cdots\Big)  
\ee
to also include the summation over the real and imaginary part of the original complex $m^{ij}_\ell$ (hence the notation $X'$), which does not really incur any computation burden for our purposes, we get the sum rules
\be
c_{ijkl}^{m,n} = \< C_{ijkl}^{m,n}\> \equiv \<   \[m_\ell^{ij} m_\ell^{kl} +(-1)^m  m_\ell^{il} m_\ell^{kj}     \] \f{C_\ell^{m,n}}{\mu^{m+n+1}}  \>  ,~~m\geq 2,n\geq 0   ,
\ee
where now $m^{ij}_\ell(\mu)$ are real numbers for any $\mu$ and $\ell$. For the bi-scalar example to be explicitly considered later, we will focus on the time reversal invariant case.

\subsection{$st$ null constraints}

The sum rules (\ref{csumrules}) come from directly expanding the $su$ symmetric dispersion relation. However, the amplitude actually contains $stu$ triple crossing symmetries. (For scattering between two identical particles, they are really symmetries, the amplitude being invariant under permutations of $stu$; for the case of multi-field scattering, they probably could be more appropriately called crossing relations, as crossing generally links different amplitudes.) Additional set of sum rules, which will be referred to the $st$ null constraints, can be extracted by imposing the $st$ crossing symmetry on the $su$ symmetric dispersion relation \cite{Tolley:2020gtv, Caron-Huot:2020cmc}.

Having obtained the sum rules (\ref{csumrules}), the easiest way to obtain the null constraints is to use the $st$ symmetry to impose null equalities on $c_{ijkl}^{m,n}$, and by replacing $c_{ijkl}^{m,n}$ with dispersive integrals via \eref{csumrules}, we also get null constraints in terms of summation over the UV states. To this end, we expand the amplitude in $s$ and $t$
\begin{equation}
\label{Bexpandion}
    B_{ijkl}(s,t)=\sum_{p,q} a_{ijkl}^{p,q}s^p t^q   .
\end{equation}
Note that the expansion coefficients $a_{ijkl}^{m,n}$ are different from $c_{ijkl}^{m,n}$, the latter being the expansion coefficients of $v=s+\f{t}2$ and $t$. For scalars, we have the $st$ crossing symmetry
\begin{equation}
\label{stcosssym}
    B_{ijkl}(s,t)=B_{ikjl}(t,s)   .
\end{equation}
(For massless fields with spin, we will still have this crossing symmetry, but the partial wave expansion will be more complicated; for example, in the helicity formalism, the Wigner $d$-matrices should be used instead.) Plugging \eref{Bexpandion} into \eref{stcosssym}, we see that the null constraints are simply 
\be
a_{ijkl}^{p,q}=a_{ikjl}^{q,p}   .
\ee
To convert to the relations between $c_{ijkl}^{m,n}$, the $s,t$ expansion can be matched to the $v,t$ expansion $\tilde B_{ijkl}(v,t)=\sum_{m,n\geq 0}c_{ijkl}^{m,n}\(s+\f{t}2\)^m t^n$. Note that here the $v$ expansion starts from $m=0$, and we have defined some new coefficients $c_{ijkl}^{0,n}$ and $c_{ijkl}^{1,n}$, which are just the Taylor expansion coefficients of $\td a^{(0)}_{ijkl}(t)$ and $a^{(1)}_{ijkl}(t)$
\be
\td a^{(0)}_{ijkl}(t)  =  \sum_{n\geq 0} c^{0,n}_{ijkl} t^n,~~~~a^{(1)}_{ijkl}(t)  =  \sum_{n\geq 0} c^{1,n}_{ijkl} t^{n}  .
\ee 
By binomially expanding $(s+\f{t}2)^m$ and relabeling the summations, we get null constraints in terms of the $su$ symmetric coefficients $c_{ijkl}^{m,n}$: 
\be
\label{nullinC}
{\bar{n}}_{ijkl}^{p,q} =a_{ijkl}^{p,q}-a_{ikjl}^{q,p} =\sum_{a=p}^{p+q} \f{\Gi(a+1)c_{ijkl}^{a,p+q-a}}{2^{a-p}\Gi(p+1)\Gi(a-p+1)} -\sum_{b=q}^{p+q}  \f{\Gi(b+1) c_{ikjl}^{b,p+q-b}}{2^{b-q}\Gi(q+1)\Gi(b-q+1)} =0   .
\ee
As mentioned above, the $st$ crossing symmetry further allows us to expand $c^{0,n}_{ijkl}$ and $c^{1,n}_{ijkl}$ in terms of the $c_{ijkl}^{m\geq2,n}$ coefficients, which is not surprising as we are using twice subtracted dispersion relation. To see this, we impose the conditions $B_{ijkl}(0,t)=B_{ikjl}(t,0)$ and $B_{ijkl}(-2t,t)=B_{ikjl}(t,-2t)$, which are respectively given by
\bal
\label{B0tBt0}
 \td a^{(0)}_{ijkl}(t) + a^{(1)}_{ijkl}(t) \f{t}2 + \sum_{m\geq 2,n\geq 0} c^{m,n}_{ijkl}\(\f{t}2\)^m t^n &= \td a^{(0)}_{ikjl}(0) + a^{(1)}_{ikjl}(0)t + \sum_{m\geq 2} c^{m,0}_{ikjl}t^m    ,
 \\
 \td a^{(0)}_{ijkl}(t) - a^{(1)}_{ijkl}(t) \f{3t}2 + \sum_{m\geq 2,n\geq 0} c^{m,n}_{ijkl}\(-\f{3t}2\)^m t^n &= \td a^{(0)}_{ikjl}(-2t)     .
\eal
Combining these two equations allows us to eliminate $a^{(1)}_{ijkl}(t)$ on the left hand side and get an equation going like 
\be
\label{a0a1mid}
4\td a^{(0)}_{ijkl}(t)-\td a^{(0)}_{ikjl}(-2t) =(\cdots)   .
\ee 
Shifting $t\to -2t$, swapping $j\leftrightarrow k$ and dividing both sides by $1/4$ for \eref{a0a1mid}, we get another equation going like $\td a^{(0)}_{ikjl}(-2t)-\f14\td a^{(0)}_{ijkl}(4t) =(\cdots)$. Adding the two equations, we get $4\td a^{(0)}_{ijkl}(t)-\f14 \td a^{(0)}_{ijkl}(4t) =(\cdots)$, which can be used to solve for $c^{0,n}_{ijkl}$. After this, we can substitute the result of $\td a^{(0)}_{ijkl}(t)$ back into \eref{B0tBt0}, which then allows us to solve for $c^{1,n}_{ijkl}$. The coefficients $c^{0,0}_{ijkl}$, $c^{0,1}_{ijkl}$, $c^{1,0}_{ijkl}$, $c^{1,1}_{ijkl}$ and $c^{0,2}_{ijkl}$ will not affect non-trivial null constraints. The first few $c^{0,n}_{ijkl}$ and $c^{1,n}_{ijkl}$ that do go into the non-trivial null constraints are explicitly given by 
\bal
c^{0,3}_{ijkl}=&\f14c^{2,1}_{ijkl}+\f14c^{3,0}_{ijkl}-\f12c_{ikjl}^{2,1}+\f14c^{3,0}_{ikjl}\\c^{1,2}_{ijkl}=&-c^{2,1}_{ijkl}-\f34c^{3,0}_{ijkl}+c^{2,1}_{ikjl}+\f32c^{3,0}_{ikjl}\\
c^{0,4}_{ijkl}=&-\f18c_{ijkl}^{3,1}-\f3{16}c_{ijkl}^{4,0}+\f14c_{ikjl}^{2,2}-\f18c_{ikjl}^{3,1}+\f38c_{ikjl}^{4,0}\\
c^{1,3}_{ijkl}=&\f34c_{ijkl}^{3,1}+c_{ijkl}^{4,0}-c_{ikjl}^{2,2}-\f12c_{ikjl}^{3,1}+\f12c_{ikjl}^{4,0}\\
c^{0,5}_{ijkl}=&\f1{16}c_{ijkl}^{4,1}+\f18c_{ijkl}^{5,0}-\f18c_{ikjl}^{2,3}+\f1{16}c_{ikjl}^{3,2}-\f3{16}c_{ikjl}^{4,1}+\f7{32}c_{ikjl}^{5,0}\\
c^{1,4}_{ijkl}=&-\f1{2}c_{ijkl}^{4,1}-\f{15}{16}c_{ijkl}^{5,0}+\f34c_{ikjl}^{2,3}+\f1{8}c_{ikjl}^{3,2}+\f1{8}c_{ikjl}^{4,1}+\f{15}{16}c_{ikjl}^{5,0}   .
\eal
Substituting the above relations into \eref{nullinC}, we get null constraints in terms of $c_{ijkl}^{m\geq2,n}$
\be
\bar n_{ijkl}^{p,q}=(a_{ijkl}^{p,q}-a_{ikjl}^{q,p})|_{c^{0,n},\,c^{1,n}\to c^{m\geq 2,n}}=0   ,
\ee
the first few of which are listed in Appendix \ref{sec:EF}. We find the first nontrivial null constraint comes in at $p+q=4$, which is similar to the single scalar case and related to the fact that the dispersion relation needs twice subtraction. Not surprisingly, there are more null constraints in the multi-field case than in the single field case at each order.

In practice, we can further impose $jl$ symmetry for $\bar n_{ijkl}^{p,q}=0$ and have
\be
\label{nullconsts0}
n^{ijkl}=\bar n_{i(j|k|l)}^{p,q}=0   ,
\ee 
which is simply a recognition of the $su$ crossing symmetry of the dispersion relation. To see this, note that, performing $jl$ crossing on \eref{stcosssym}, we get $B_{ilkj}(u,t)=B_{iklj}(t,u)$, and then relabeling $u$ to $s$ gives $B_{ilkj}(s,t)=B_{iklj}(t,s)$, which is just $B_{ijkl}(s,t)=B_{ikjl}(t,s)$ with $jl$ swapped. This means that $su$ crossing symmetry implies the null constraints $B_{ijkl}(s,t)=B_{ikjl}(t,s)$ is symmetric in exchanging $jl$. As $su$ crossing symmetry is built-in in the $su$ dispersion relation, imposing $jl$ symmetry for $\bar n_{ijkl}^{p,q}=0$ does not incur any loss in any meaningful null constraints. This is also consistent with what we find in our numerical SDP studies later. The first few nontrivial and independent null constraints from $n^{ijkl}=0$ are given by:
\bal
\label{nConstraintsStart}
n_{ijkl}^{1,3}&=c_{i(j|k|l)}^{2,2}+\f34 c_{i(j|k|l)}^{3,1}+\f12 c_{i(j|k|l)}^{4,0}-c_{ik(jl)}^{3,1}-2c_{ik(jl)}^{4,0}=0
\\
n_{ijkl}^{2,2}&=c_{i(j|k|l)}^{2,2}+\f32 (c_{i(j|k|l)}^{3,1}+ c_{i(j|k|l)}^{4,0})- c_{ik(jl)}^{2,2}-\f32(c_{ik(jl)}^{3,1}+c_{ik(jl)}^{4,0})=0
\\
n_{ijkl}^{1,4}&=c_{i(j|k|l)}^{2,3}+\f34c_{i(j|k|l)}^{3,2}+\f12c_{i(j|k|l)}^{4,1}+\f5{16}c_{i(j|k|l)}^{5,0}-c_{ik(jl)}^{4,1}-\f52c_{ik(jl)}^{5,0}=0
\\
n_{ijkl}^{2,3}&=c_{i(j|k|l)}^{2,3}+\f32c_{i(j|k|l)}^{3,2}+\f32c_{i(j|k|l)}^{4,1}+\f5{4}c_{i(j|k|l)}^{5,0}-c_{ik(jl)}^{3,2}-2c_{ik(jl)}^{4,1}-\f52c_{ik(jl)}^{5,0}=0
\\
n_{ijkl}^{1,5}&=c_{i(j|k|l)}^{2,4}+\f34c_{i(j|k|l)}^{3,3}+\f12c_{i(j|k|l)}^{4,2}+\f5{16}c_{i(j|k|l)}^{5,1}+\f3{16}c_{i(j|k|l)}^{6,0} -c_{ik(jl)}^{5,1}-3c_{ik(jl)}^{6,0}=0
\\
n_{ijkl}^{2,4}&=c_{i(j|k|l)}^{2,4}+\f32c_{i(j|k|l)}^{3,3}+\f32c_{i(j|k|l)}^{4,2}+\f5{4}c_{i(j|k|l)}^{5,1}+\f{15}{16}c_{i(j|k|l)}^{6,0}-c_{ik(jl)}^{4,2}- \f52c_{ik(jl)}^{5,1}-\f{15}4c_{ik(jl)}^{6,0}=0
\\
n_{ijkl}^{3,3}&=c_{i(j|k|l)}^{3,3}+ 2c_{i(j|k|l)}^{4,2}+\f5{2}(c_{i(j|k|l)}^{5,1}+c_{i(j|k|l)}^{6,0})-c_{ik(jl)}^{3,3}- 2c_{ik(jl)}^{4,2}-\f52(c_{ik(jl)}^{5,1}+c_{ik(jl)}^{6,0})=0   .
\label{nConstraintsEnd}
\eal
These null constraints restrict the viable Wilson coefficients to be in a linear subspace of the original linear space spanned by $c_{ijkl}^{m,n}$. Put it another way, the viable Wilson coefficients live in the null space of the homogeneous linear system of the null constraints.

There are at least two ways to impose these null constraints. The direct way is to first extract the (potential) positivity region in the parameter space of $c_{ijkl}^{m,n}$ from the $su$ symmetric dispersion relation, which can be done with the SDP method that will be described momentarily, and then use the null constraints (\ref{nullconsts0}) to slice out the linear subspace that satisfies the $st$ crossing symmetry, as emphasized by  \cite{Chiang:2021ziz} for the case of a single scalar. That is, the fully crossing symmetric positivity bounds are given by the intersection between the $su$ symmetric positivity bounds and the null space of the $st$ null constraints. In practice, the first step of this method can often be cumbersome to implement with the SDP method, as one may need to extract the $su$ positivity bounds for quite a few Wilson coefficients before one can use a number of the $st$ null constraints to reduce to the fully crossing symmetric subspace --- there are typically quite a few coefficients in these null constraints, especially in a multi-field EFT.  The alternative way is to first use the sum rules (\ref{csumrules}) to convert the null constraints (\ref{nullconsts0}) to a set of null dispersive integrals, and then linearly add these null dispersive integrals to the sum rules (\ref{csumrules}), which again can be turned into an SDP, as we shall see shortly. The upshot of this method is that one can efficiently constrain some selected Wilson coefficients that are of concern for a particular problem. This is the approach we will take for the examples in this paper. 

To cast the null constraints in terms of the dispersive integrals, we substitute the sum rules (\ref{csumrules}) into the above constraints and end up with null constraints of the form
\bal
\bigg\langle \[m_\ell^{ij} (m_\ell^{kl})^*   +   m_\ell^{il} (m_\ell^{kj})^* \] \f{E^{+}_{p,q}}{\mu^{p+q+1}} +  \[m_\ell^{ij} (m_\ell^{kl})^* -  m_\ell^{il} (m_\ell^{kj})^* \]  \f{E^{-}_{p,q}}{\mu^{p+q+1}}   ~~ &
\nn
+ \[m_\ell^{ik} (m_\ell^{jl})^*   +   m_\ell^{il} (m_\ell^{jk})^* \] \f{F^{+}_{p,q}}{\mu^{p+q+1}} +  \[m_\ell^{ik} (m_\ell^{jl})^* -  m_\ell^{il} (m_\ell^{jk})^* \]  \f{F^{-}_{p,q}}{\mu^{p+q+1}}   \bigg\rangle 
&=0   ,
\label{rawNull}
\eal
where $E^{\pm}_{p,q}$ and $F^{\pm}_{p,q}$ are polynomials of $\ell$ and the first few of them that appeared in (\ref{nConstraintsStart}-\ref{nConstraintsEnd}) are given in Appendix \ref{sec:EF} for a quick reference. In the next subsection, we will formulate the positivity bounds as the outcome of a tractable SDP solvable with {\nb SDPB}. In order to do that, as will become clearly shortly, viable null constraints should be put in a form with only $m_\ell^{ij} (m_\ell^{kl})^*$ in dispersive integral. This can be achieved by contracting the above sum rule with a general tensor $\mc{N}_{p,q}^{ijkl}$ and we get
\bal
\label{nullcons}
\< \hat N^{({\cal N})}_{p,q}\> & \equiv \sum_{i,j,k,l}\bigg\langle   \f{m_\ell^{ij} (m_\ell^{kl})^*}{\mu^{p+q+1}}   N^{({\cal N})ijkl}_{p,q}  \bigg\rangle
\\
&\equiv \sum_{i,j,k,l}\bigg\langle   \f{m_\ell^{ij} (m_\ell^{kl})^*}{\mu^{p+q+1}}  \Big[  \mc{N}_{p,q}^{i(j|k|l)}  E^{+}_{p,q} + \mc{N}_{p,q}^{i[j|k|l]}  E^{-}_{p,q} 
 +  \mc{N}_{p,q}^{ik(jl)}  F^{+}_{p,q} +  \mc{N}_{p,q}^{ik[jl]}   F^{-}_{p,q}  \Big]  \bigg\rangle =0   .
\eal
That is, to get all available null constraints, we should survey all possible forms for the constant tensor $\mc{N}_{p,q}^{ijkl}$. This means that in a generic multi-field EFT there can be many more null constraints than in the single scalar case. If the external particles $i,j,k,l$ are endowed with some symmetries, however, the form of $\mc{N}^{ijkl}$ may be more restricted. Note that as expected the constraints $\<|m^{11}_\ell |^2 \big(E^{+}_{p,q}+F^{+}_{p,q}\big)/\mu^{p+q+1}\>=0$, which can be obtained by choosing ${\cal N}_{p,q}^{ijkl}=\delta^{i}_1\delta^j_1\delta^k_1\delta^l_1$, are exactly the null constraints from identical scalar scattering.

\section{Multi-field positivity bounds with full crossing}

\label{sec:sdp}

With the sum rules (\ref{csumrules}) and the null constraints (\ref{nullcons}) established, we are now ready to formulate our problem as a convex optimization program to obtain the optimal positivity bounds. For the single scalar case, the optimal bounds can be obtained via linear programing. For the multi-field case, full-blown semi-definite programing with one continuous decision variable $\mu$ is generally needed, which however is still directly solvable via the powerful package {\nb SDPB}, by now a standard tool in CFT bootstrap \cite{Poland:2018epd}.

\subsection{Semi-definite program}

In the sum rules for $c^{m,n}_{ijkl}$, \ie \eref{csumrules}, we see that the expression inside the angle bracket splits into two parts: the part with $m_\ell^{ij} (m_\ell^{kl})^* +(-1)^m m_\ell^{il} (m_\ell^{kj})^*$, which depends on the UV theory and is mostly unknown, and the part with ${C_\ell^{m,n}}/{\mu^{m+n+1}}$, where $C_\ell^{m,n}$ is a known polynomial of $\ell$ and $\mu$ is the unknown scale of the UV modes. A naive approach would be to formulate an optimization problem with all of $\mu$ and $m_\ell^{ij}(\mu)$ as decision variables, which, however, is intractable as, even if we cut $\ell$ at some finite integer $\ell_{\rm  max}$ and approximate the continuous $\mu$ with $N_{\mu}$ discrete points, $m_\ell^{ij}(\mu)$ still represents about $N^2\ell_{\rm max}N_\mu$ continuous decision variables. 

To proceed, we first note that there is a sign difference in $m_\ell^{ij} (m_\ell^{kl})^* +(-1)^m m_\ell^{il} (m_\ell^{kj})^*$ between sum rules $c_{ijkl}^{m,n}$ with even $m$ and odd $m$
\bal
c_{ijkl}^{2h,n} &= \< C_{ijkl}^{2h,n}\> \equiv \<   \[m_\ell^{ij} (m_\ell^{kl})^* + m_\ell^{il} (m_\ell^{kj})^*     \] \f{C_\ell^{2h,n}}{\mu^{2h+n+1}}  \>   ,
\\
c_{ijkl}^{2h+1,n} &= \< C_{ijkl}^{2h+1,n}\> \equiv \<   \[m_\ell^{ij} (m_\ell^{kl})^* - m_\ell^{il} ( m_\ell^{kj})^*     \] \f{C_\ell^{2h+1,n}}{\mu^{2h+n+2}}  \>   .
\eal
So, to construct an SDP to constrain multiple $c_{ijkl}^{m,n}$, we can introduce a decision variable tensor $\mc{Q}^{ijkl}_{m,n}$ that is symmetric for even $m$ and anti-symmetric for odd $m$ when exchanging $j$ and $l$. That is, if there exists a constant tensor $\mc{Q}^{ijkl}_{m,n}$ with the symmetries
\be
\mc{Q}^{ijkl}_{2h,n} =   \mc{Q}^{ilkj}_{2h,n},~~~~~\mc{Q}^{ijkl}_{2h+1,n} =  - \mc{Q}^{ilkj}_{2h+1,n}   
\ee
such that the following conditions 
\bal
\label{Qconedual}
\sum_{i,j,k,l;m,n} \mc{Q}_{m,n}^{ijkl} C_{ijkl}^{m,n} &= \sum_{i,j,k,l;m,n} \mc{Q}_{m,n}^{ijkl}     \[m_\ell^{ij} (m_\ell^{kl})^* + (-1)^m  m_\ell^{il} (m_\ell^{kj})^*     \] \f{C_\ell^{m,n}}{\mu^{m+n+1}} 
\\
 &=\sum_{i,j,k,l}  m_\ell^{ij}  \[ \sum_{m,n} {\mc{Q}^{ijkl}_{m,n}}  \f{C_\ell^{m,n}}{\mu^{m+n+1}}  \] (m_\ell^{kl})^* \geq 0  ,~~~~~{\rm for~all}~ \mu,\ell, m_\ell^{ij}(\mu) ,
 \label{QCconstraint}
\eal
are satisfied, then we have {\it a positivity bound}
\be
\sum_{i,j,k,l;m,n} \mc{Q}^{ijkl}_{m,n}   c_{ijkl}^{m,n} = \< \sum_{i,j,k,l;m,n} \mc{Q}^{ijkl}_{m,n}   C_{ijkl}^{m,n}\> \geq 0   .
\ee
As it stands, the constraint conditions (\ref{QCconstraint}) are rather complicated, as it still contains $m_\ell^{ij}(\mu)$. However, it is easy to see that, if we view $m_\ell^{ij}$ as a vector (viewing $ij$ as one index), the inequality (\ref{QCconstraint}) is simply a quadratic form, and the condition simply means that $\sum_{m,n}{\mc{Q}^{ijkl}_{m,n}} C_\ell^{m,n}/{\mu^{m+n+1}}$ as a Hermitian matrix with indices $ij$ and $kl$ is positive semi-definite positive. Thus, whether a set of Wilson coefficients satisfy the positivity bounds can be determined by checking whether the objective of the following SDP, $\sum_{i,j,k,l;m,n} \mc{Q}^{ijkl}_{m,n}  c_{ijkl}^{m,n}$, has a non-negative minimum: given a set of Wilson coefficients $c_{ijkl}^{m,n}$,
\bal
{\rm minimize}& ~~~  \sum_{i,j,k,l;m,n} \mc{Q}^{ijkl}_{m,n}  c_{ijkl}^{m,n}
\\
{\rm subject~to}& ~~~ 
\sum_{m,n} \mc{Q}^{ijkl}_{m,n}   \f{C_\ell^{m,n}}{\mu^{m+n+1}}  \succeq 0
~~~{\rm for~all}~ \mu,\ell  ,
\label{Qspecon}
\eal
where $X^{ij,kl}\succeq 0$ denotes that the matrix $X$ with indices $ij$ and $kl$ is positive semi-definite. Now, this program only has one continuous variable $\mu$, which as we shall see in the next subsection is manageable. The collection of all feasible sets of Wilson coefficients $c_{ijkl}^{m,n}$ to the SDP forms the positivity region in the $c_{ijkl}^{m,n}$ parameter space. The boundaries of the positivity region, \ie the positivity bounds, are where the objective of the SDP vanishes
\be
\sum_{m,n}\mc{Q}_{m,n} \cdot c^{m,n}\geq 0   ,
\ee 
where, to avoid clutter in the indices, we have suppressed the summation over $ijkl$ and adopted the shorthand notation
\be
X\cdot c \equiv \sum_{i,j,k,l} X^{ijkl} c_{ijkl}, ~~~{\rm and}~~~  m.X.m^* \equiv \sum_{i,j,k,l} m^{ij}X^{ijkl}(m^{kl})^*   .
\ee

Let us recapitulate from the point view of convex geometry. We can view $c_{ijkl}^{m,n}$ as elements of a convex cone, which is generated by conical combinations of $[m_\ell^{ij} (m_\ell^{kl})^* + (-1)^m m_\ell^{il} (m_\ell^{kj})^* ] {C_\ell^{m,n}}/{\mu^{m+n+1}}$. The positivity bounds $\sum_{m,n}\mc{Q}_{m,n} \cdot c^{m,n}\geq 0$ define a dual cone whose elements are $\mc{Q}^{ijkl}_{m,n}$. The optimal positivity bounds are given by the boundaries of the $c_{ijkl}^{m,n}$ cone or the extremal rays of $\mc{Q}^{ijkl}_{m,n}$. However, from \eref{Qspecon}, we see that it is $\sum_{m,n} \mc{Q}^{ijkl}_{m,n} {C_\ell^{m,n}}/{\mu^{m+n+1}}$ that lives in a spectrahedron, the viable space of the decision variables of an SDP. So, at technical level, we can view the (partial) ``spectral tensor'' $m_\ell^{ij} (m_\ell^{kl})^*$ as forming a convex cone whose elements are indexed by $i,j,k,l$ and whose dual cone is the spectrahedron $\sum_{m,n} \mc{Q}^{ijkl}_{m,n} {C_\ell^{m,n}}/{\mu^{m+n+1}}$.

In the SDPs above to obtain the positivity bounds, we have only made use of the $su$ symmetric dispersion relation and have yet to take into account the $st$ crossing symmetry. As we mentioned previously, one way to impose the $st$ crossing symmetry is to intersect the positivity bound region with the $st$ symmetric null constraints (\ref{nullconsts0}), if we have computed bounds for sufficiently many Wilson coefficients. In the following, however, we will take an alternative approach, to impose null constraints with dispersive integrals (\ref{nullcons}). As we often want to know positive bounds for a few specific Wilson coefficients, the latter approach is usually more convenient for this purpose.

The essential idea is that we can add the null constraints with $\hat N^{({\cal N})}_{p,q}$ from the previous section to the SDP above. Since the null constraints are null within the average $\< \cdots \>$, the objective function of the SDP is unchanged (still $\sum_{m,n}  \mc{Q}_{m,n} \cdot c^{m,n}$), but the null constraints do alter the linear matrix inequalities of the SDP via adding terms with $\hat N^{({\cal N})}_{p,q}$, alongside with new decision variables, as this is done without imposing the average $\< \cdots \>$. Thanks to the fact that we can add the null constraints with decision variables of opposite signs, this new SDP will bound the Wilson coefficients from different directions, leading to the EFT couplings to be constrained in an enclosed convex region. More concretely, if there exist  constant tensor $\mc{N}^{ijkl}_{p,q}$ and constant tensor $\mc{Q}^{ijkl}_{m,n}$ with the symmetries
\bal
\mc{Q}^{ijkl}_{2h,n} =   \mc{Q}^{ilkj}_{2h,n}, ~~~\mc{Q}^{ijkl}_{2h+1,n} =  - \mc{Q}^{ilkj}_{2h+1,n}
\eal
such that
\bal
 \label{QNconstraints}
&~~~\sum_{m,n} \mc{Q}_{m,n} \cdot C^{m,n}  +\sum_{{\cal N},p,q}  {\hat N}^{{(\cal N)}}_{p,q} 
 ,~~~~~{\rm for~all}~ \mu,\ell, m_\ell^{ij} ,
\nonumber \\
 &= m_\ell . \[ \sum_{m,n} {\mc{Q}_{m,n}}  \f{C_\ell^{m,n}}{\mu^{m+n+1}}  + \sum_{{\cal N},p,q} \f{N^{({\cal N})}_{p,q}}{\mu^{p+q+1}}    \] . (m_\ell)^* \geq 0     ,
\eal
then we have a positivity bound
\be
\sum_{i,j,k,l;m,n} \mc{Q}^{ijkl}_{m,n}   c_{ijkl}^{m,n} = \< \sum_{i,j,k,l;m,n} \mc{Q}^{ijkl}_{m,n}   C_{ijkl}^{m,n}\> \geq 0   .
\ee
In the language of convex optimization, the linear matrix inequalities now construct a cone in a larger space with both the $c_{ijkl}^{m,n}$ and $n_{ijkl}^{p,q}$ coefficients, but in the objective function it is projected down to the subspace of the $c_{ijkl}^{m,n}$ coefficients. Clearly, with the null constraints included, whether a set of Wilson coefficients satisfy the positivity bounds can be determined by checking whether the objective of the following stronger SDP has a non-negative minimum: given a set of Wilson coefficients $c_{ijkl}^{m,n}$,
\bal
{\rm minimize}& ~~~  \sum_{m,n} \mc{Q}_{m,n}  \cdot c^{m,n} 
\label{sdpmain}
\\
{\rm subject~to}& ~~~ 
\sum_{m,n} \mc{Q}_{m,n}   \f{C_\ell^{m,n}}{\mu^{m+n+1}}   + \sum_{{\cal N},p,q} \f{N^{({\cal N})}_{p,q}}{\mu^{p+q+1}}   \succeq 0
~~~{\rm for~all}~ \mu,\ell    ,
\label{lminull}
\eal
to have a non-negative minimum. We emphasize that we have suppressed the $ijkl$ indices for $\mc{Q}_{m,n}$ and $N^{({\cal N})}_{p,q} $, which are considered as matrices when evaluating the positive semi-definiteness ``$\succeq 0$''. The collection of all feasible sets of Wilson coefficients $c_{ijkl}^{m,n}$ to the SDP form the positivity region in the $c_{ijkl}^{m,n}$ space and the boundaries of the positivity region are again where the objective $\mc{Q}_{m,n} \cdot c^{m,n}$ vanishes.

We want to emphasize that, by adding the null constraints to the SDP, we have utilized more information of the amplitude, \ie the triple crossing symmetry, in extracting the constraints on the Wilson coefficients. The importance of these extra terms in the linear matrix inequalities (\ref{lminull}) is that it generally allows us to constrain the coefficients from opposite directions. This is possible because the contributions from the null constraints can dominate the left hand side of the linear matrix inequalities (\ref{lminull}), which makes it possible for $\mc{Q}_{m,n}^{ijkl}$ to have both signs \cite{Tolley:2020gtv, Caron-Huot:2020cmc}. More concretely, suppose the SDP (\ref{sdpmain} and \ref{lminull}) gives rise to a bound of the form $\sum_{m,n} \mc{Q}_{m,n}  \cdot c^{m,n}\geq 0$ for a given set of decision variables $\mc{Q}_{m,n}^{ijkl}$ and $\mc{N}^{ijkl}_{p,q}$; thanks to the null constraints, the SDP may also be feasible for $\bar {\mc{Q}}_{m,n}^{ijkl}= -\mc{Q}_{m,n}^{ijkl}$, which will be facilitated by a different set of $\bar{\mc{N}}^{ijkl}_{p,q}$, and thus the SDP gives rise to another bound of the form $ \sum_{m,n} \mc{Q}_{m,n}  \cdot c^{m,n}\leq 0$.

\subsection{Implementation}
\label{sec:impl}

As we have seen, generally, our SDP problem is defined in the complex domain with complex linear matrix inequality constraints. Efficient SDP algorithms are usually implemented in the real domain. The reason is that a complex SDP problem can be easily transformed to a real SDP problem by using the fact that a Hermitian matrix $H$ is positive semi-definite if and only if a doubly enlarged matrix constructed from ${\rm Re} H$ and ${\rm Im} H$ is positive semi-definite
\be
H \succeq 0 ~~~\Longleftrightarrow ~~~\left[\begin{array}{cc}
{\rm Re} H & -{\rm Im} H \\
{\rm Im} H & {\rm Re} H
\end{array}\right] \succeq 0   ,
\ee
where ${\rm Re} H$/${\rm Im} H$ is the real/imaginary part of the complex Hermitian matrix $H$. However, as we mentioned previously, in the following, we will focus on exemplary theories with time reversal invariance, for which, according to the argument around \eref{mmRealNot}, $m_\ell^{ij}(\mu)$ can be taken as real matrices
\be
m^{ij}_\ell (m^{kl}_\ell)^* ~~\to~~ m^{ij}_\ell  m^{kl}_\ell,~~~\text{time reversal invariant}   .
\ee
With this simplification, it is already an SDP in the real domain. The special feature of this SDP is that it has a continuous decision variable $\mu$ that is in the denominators and goes from $\Lambda^2$ to $\infty$. This is hardly a problem, as it can be easily transformed to a polynomial matrix program, solvable by {\nb SDPB} \cite{Simmons-Duffin:2015qma}. 
 
For readers unfamiliar with {\nb SDPB}, this package is able to efficiently solve the following polynomial matrix program (by transforming it into a standard SDP): given $A+1$ real numbers $b_a$ and $J(A+1)$ real symmetric matrices of dimension $n_{j}\times n_{j}$
\be
M_{j}^{a}(x)=\left(\begin{array}{ccc}
P_{j, 11}^{a}(x) & \ldots & P_{j, 1 n_{j}}^{a}(x) \\
\vdots & \ddots & \vdots \\
P_{j, n_{j} 1}^{a}(x) & \ldots & P_{j, n_{j} n_{j}}^{a}(x)
\end{array}\right) ,~~P_{j, uv}^{a}(x)=P_{j, vu}^{a}(x),~~~ a=0,1,...,A,~~~j=1,2,...,J    ,
\ee
with each element $P_{j,uv}^{a}(x)$ being a polynomial of $x$, the package can survey all possible real numbers $y_a$ to
\bal
\text { maximize }&~~~ b_0+\sum^{A}_{a=1} b_a y_a\\
\text { such that }&~~~ M_{j}^{0}(x)+\sum_{a=1}^{A} y_{a} M_{j}^{a}(x) \succeq 0 \quad \text { for all } x \geq 0 \text { and } 1 \leq j \leq  J   .
\eal
Note that here the linear matrix inequality conditions are required to satisfy for $1 \leq j \leq  J$ and for all continuous $x\geq 0$.  

To convert our SDP problem (\ref{sdpmain}) into the standard form above, we can choose a basis $\mc{Q}^{ijkl}_{(a)}$ for the $\mc{Q}^{ijkl}_{m,n}$. The form of $\mc{Q}_{(a)}$ can be mapped from the symmetries of $c^{m,n}_{ijkl}$, and specifically is restricted by crossing symmetries $\mc{Q}_{m,n}^{ijkl}=(-1)^m \mc{Q}_{m,n}^{ilkj}=(-1)^m \mc{Q}_{m,n}^{kjil}=\mc{Q}_{m,n}^{jilk}$ and as well as the internal symmetries of the theory, which will be further explained with an example shortly. Other choices of $\mc{Q}_{m,n}^{ijkl}$ are simply redundant. In an appropriate basis, we then have 
\bal
\mc{Q}^{ijkl}_{m,n}=\sum_{a} y^a_{m,n}\mc{Q}^{ijkl}_{(a)}   ,
\eal
where $\{y^a_{m,n}\}$ are a set of constants. Similarly, the form of $\mc{N}^{ijkl}_{p,q}$ can be derived from the symmetries of $n^{p,q}_{ijkl}$, specifically the crossing symmetries $\mc{N}_{p,q}^{ijkl}=\mc{N}_{p,q}^{ilkj}=\mc{N}_{p,q}^{kjil}=\mc{N}_{m,n}^{jilk}$ and again the internal symmetries of the theory. Expanded in an appropriate set of basis, we have
\bal
\mc{N}^{ijkl}_{p,q}=\sum_{b} z^b_{p,q}\mc{N}^{ijkl}_{(b)}   .
\eal
So, after these, the decision variables of the SDP are $y^a_{m,n}$ and $z^b_{m,n}$, together with the continuous variable $\mu$. Since the UV state mass scale $\mu$ ranges from $\Lambda^2$ to $\infty$, we can define a dimensionless variable
\be
x=\f{\mu}{\Lambda^2} -1  ,
\ee
which takes values from 0 to $\infty$. Multiplying by a common factor of $\mu^{M_{\rm M}}$ with $M_{\rm M}$ being the greatest power of $1/\mu$ in (\ref{sdpmain}) for a given problem, the linear matrix inequality (\ref{lminull}) becomes
\bal
\sum_{m,n,a} y^a_{m,n}\mc{Q}_{(a)}  C_\ell^{m,n} {[\Lambda^2(1+x)]^{M_{\rm M}-m-n-1}}
+\sum_{p,q,b}  z^b_{p,q}   {N^{(b)}_{p,q}}{[\Lambda^2(1+x)]^{M_{\rm M}-p-q-1}}   \succeq 0   ,
\eal
where we have defined $N^{(b)}_{p,q}\equiv N^{({\cal N})}_{p,q}|_{{\cal N}^{ijkl}_{p,q}\to{\cal N}^{ijkl}_{(b)}}$. In the following, the coefficient components projected to the basis matrices will be denoted as 
\be
\label{caQa}
c_{(a)}^{m,n}= c^{m,n}\cdot\mc{Q}_{(a)}   .
\ee
Using these notations, our SDP can be formulated as follows: a positivity bound is found if the following polynomial matrix program can 
\bal
\text{minimize} &~~~  \sum_{a,m,n}y^a_{m,n} c^{(a)}_{m,n}
\\
\text{subject to} &~~~
\sum_{m,n,a} y^a_{m,n}\mc{Q}_{(a)}  C_\ell^{m,n} {[\Lambda^2(1+x)]^{M_{\rm M}-m-n-1}}
\nn
&~~~~~~~~
 +\sum_{p,q,b}  z^b_{p,q}   {N^{(b)}_{p,q}}{[\Lambda^2(1+x)]^{M_{\rm M}-p-q-1}}   \succeq 0
~~~~~{\rm for~all}~\ell,~\text{and}~x\geq 0    ,
\label{sdpPractical}
\eal
and the minimum of $\sum_{a,m,n}y^a_{m,n} c^{(a)}_{m,n}$ is semi-positive, with zero being a boundary point of the positivity bounds. Practically, of course, we can only include a finite number of null constraints and partial waves, and one should include sufficient numbers of them so that the results converge to the required accuracy. We emphasize that the strength of using the null constraints in the dispersive integral form is that we can selectively constrain a small number of Wilson coefficients $c^{m,n}_{ijkl}$ that are of concern in a particular problem.

\subsection{Upper bounds of the $s^2$ coefficients}

\label{sec:s2upper}

Using the upper bound of partial wave unitarity and the first null constraint, Ref \cite{Caron-Huot:2020cmc} was able to derive an upper bound for the Wilson coefficient of identical scalar scattering $c^{2,0}_{iiii}$. In a multi-field theory,  we also want to derive upper bounds on the $s^2$ coefficients other than identical scalar scattering. As with the single scalar case, we will see later that the existence of the upper bounds for all these coefficients is essential to conclude that all the 2-2 scattering coefficients are bounded in a finite region in a multi-field theory.

Let us first briefly review how it works for the $c^{2,0}_{iiii}$ coefficient. In our notation, after incorporating the first null constraint, we get the sum rule
\be
{c}^{2,0}_{iiii} =  \sum_{\ell} \int^\infty_{\Li^2} \d \mu\; \beta^\ai_{\ell}(\mu) {\rm Im} a^{iiiii}_\ell \( \f{1}{\mu^3}  - \alpha  \f{\td N^{\rm (1)}_{1,3}}{\mu^5} \)    ,
\ee
where $\alpha$ is an arbitrary constant, $\beta^\ai_{\ell}(\mu)\equiv {2^{4\ai+3}\pi^{\ai-1} \Gamma(\ai)} (2\ell+2\ai)C_\ell^{(\ai)}(1)/{ \mu^{\ai-\f12}}$ and $\td N^{(1)}_{1,3}=\ell^4+2 \ell^3-7 \ell^2-8 \ell$. Partial wave unitarity requires that ${\rm Im}\, a^{iiii}_\ell \geq  |a^{ii ii}_{\ell}|^2$, which means that ${\rm Im}\, a^{iiii}_\ell<1$. If we choose $\ai>0$, at sufficiently large $\ell$, $\mu$ can have a range from $\Lambda^2$ to $\Xi^2_{\ell,\ai}=\text{max}\big(\Lambda^2, (\ai \td N^{(1)}_{1,3})^{1/2})\big)$ where the integrand of the above dispersive integral is negative. If we take the upper bound of partial wave unitarity and subtract the negative part of the dispersive integral, we can get an inequality
\be
{c}^{2,0}_{iiii} \leq  \sum_{\ell}  \int^\infty_{\Xi^2_{\ell,\ai}} \d \mu\; \beta^\ai_{\ell}(\mu) \( \f{1}{\mu^3}  - \alpha  \f{\td N^{(1)}_{1,3}}{\mu^5} \)    .
\ee
Let us assume that the negative part of the dispersive integrand emerges when $\ell>\ell^*_\ai$ for a given $\ai_*$, and then the integration on the right hand side can be computed explicitly. Note that when $\ell$ increases, $\Xi^2_{\ell,\ai}$ also increases so that the sum over $\ell$ actually converges. The results depend on $\ell^*_\ai$ and $\ai_*$, which can be evaluated numerically, and it is found that the optimal upper bound is given by
\be
\label{c20upperbound0}
 {c}^{2,0}_{iiii}\leq \f{1.59(4\pi)^2}{\Lambda^4}    ,
\ee
which occurs at $\ai_*=0.025$ and $\ell^*_\ai=2$. 

In a multi-field EFT, while ${\rm Im}\, a^{ijkl}_\ell$ is still bounded for generic $ijkl$, thanks to partial wave unitarity, it may not be positive definite. So the simple inequality method above does not generically apply. However, extracting the upper bounds on ${c}^{2,0}_{ijkl}$ can be readily formulated as an optimization problem if one discretizes the integrals in the ${c}^{2,0}_{ijkl}$ sum rules and the null constraints by sampling only discrete values of $\mu_n$ \cite{Chen:2023bhu, Hong:2024fbl}:
\be
\label{upperSDP}
{c}^{2,0}_{ijkl} = \int \d \mu\, {\rm Im}\, a^{ijkl}_\ell(\mu)\, (\cdots) \to {c}^{2,0}_{ijkl} \simeq \sum_n \delta \mu\, {\rm Im}\, a^{ijkl}_\ell(\mu_n)\, (\cdots),~~~\text{(similar for null constraints)}
\ee
After the discretization, we impose partial wave unitarity conditions on ${\rm Im}\, a^{ijkl}_\ell(\mu_n)$ for a finite number of $\ell$ and all the discretized $\mu_n$. Along with the null constraints, this becomes a semi-definite program \cite{Hong:2024fbl}, or a simpler linear program if the linearized unitarity conditions are used \cite{Chen:2023bhu}, which sometimes approximates the full unitarity conditions very well. Readers are referred to \cite{Chen:2023bhu,Hong:2024fbl} for the details of the strategy and the numerical implementation. While the above inequality method is computationally cumbersome to include more null constraints, it is easy to use this discretization method to improve the bound (\ref{c20upperbound0}) with more null constraints.  Combining these upper bounds with the lower bounds, which form a salient convex cone for the $v^2$ (or $s^2$) coefficients \cite{Li:2021lpe}, the $v^2$ coefficients can be constrained to be a ``capped'' salient cone.

\section{Full crossing bounds on bi-scalar theory}

\label{sec:bi-scalar}

With the formalism established, in this section we shall illustrate how to obtain the triple crossing symmetric bounds in practice. The multi-field case is different from the single field case in a number of aspects. With multiple light fields, the number of Wilson coefficients proliferates very quickly with the increase of the number of fields. This does not present any particular difficulty in determining whether a given set of coefficients are within the positivity bounds {\it per se}. However, for illustration purposes, we shall limit ourselves to the simple case of bi-scalar theory endowed with some discrete symmetries, for which it is easier to visualize the geometric shapes of the bounds. We will start with the simpler version where the two scalars are invariant under a double $\mathbb{Z}_2$ symmetry in 4 dimensional spacetime. We will see how the $v^2t$ coefficients are bounded in a finite region, and then compute the two-sided bounds for the higher order coefficients. We then relax the symmetry slightly, to let the theory only have one $\mathbb{Z}_2$ symmetry. We find that the geometric shapes now become more complex, but still one can bound the coefficients in a small finite region.

\subsection{Bi-scalar theory with double $\mathbb{Z}_2$ symmetry}

We shall start with a simple case of two scalars with a double $\mathbb{Z}_2$ symmetry in 4D, that is, a scalar field theory invariant under two discrete transformations
\begin{itemize}
\item $\phi_1 \leftrightarrow \phi_2$
\item $\phi_1\to -\phi_1$ (the $\phi_1 \leftrightarrow \phi_2$ symmetry implies that we also have $\phi_2\to -\phi_2$)
\end{itemize}
To perform the SDP optimization, we can first find a basis for ${\cal Q}_{(a)}^{ijkl}$ and $\mc{N}^{ijkl}_{(a)}$, which are determined by crossing symmetries and internal symmetries of the theory, as mentioned in Section \ref{sec:impl}. To be precise, the forms of ${\cal Q}_{(a)}^{ijkl}$ and $\mc{N}^{ijkl}_{(b)}$ are determined by the symmetries of the Wilson coefficients $c_{ijkl}^{m,n}$. (Of course, more general ${\cal Q}_{(a)}^{ijkl}$ and $\mc{N}^{ijkl}_{(b)}$ can be used, but that is just redundancy and can significantly increase the computational costs when there are many light modes.) For our particular example, the reflection symmetry $\phi_i\to -\phi_i$ requires that ${\cal Q}_{(a)}^{iiij}$ ($i\neq j$) and its cyclic permutations vanish. The (time reversal invariant) $su$ crossing symmetry requires ${\cal Q}_{(a)}^{1122}={\rm sign}(a){\cal Q}_{(a)}^{1221}={\rm sign}(a){\cal Q}_{(a)}^{2112}={\cal Q}_{(a)}^{2211}$, where we have used the sign of integer $a$ to differentiate the basis vector for ${\cal Q}_{2h,n}^{ijkl}$ (with $a>0$) and ${\cal Q}_{2h+1,n}^{ijkl}$ (with $a<0$). Also, for negative $a$, we have $\mc{Q}^{ijij}_{(a)}=-\mc{Q}^{ijij}_{(a)}=0$ ($i$ and $j$ can be the same). The $i\leftrightarrow j$ and $k\leftrightarrow l$ crossing symmetries require ${\cal Q}_{(a)}^{1212}={\cal Q}_{(a)}^{2121}$. After these considerations, for the case of $\mathbb{Z}_2$ symmetry $\phi_1\to -\phi_1$, we find that a generic ${\cal Q}_{m,n}$ takes the form
\be
\label{Qmnform}
	{\cal Q}_{2h,n}=
\begin{blockarray}{lccccr}
   kl&11 & 22 & 12 & 21 &~~~ij\\
\begin{block}{l(cccc)r}
	&x_1 & x_3 & 0 & 0  &11\\
	&x_3 & x_2 & 0 & 0  & 22\\
	&0 & 0 & x_4 & x_3 & 12\\
	&0 & 0 & x_3 & x_4 & 21\\
\end{block}
\end{blockarray}
,~~~~~
\mc{Q}_{2h+1,n}=\begin{pmatrix}
&0&x_5&0&0\\
& x_5 &0&0&0\\
&0&0&0&-x_5\\
&0&0&-x_5&0
\end{pmatrix}    ,
\ee
{\vskip -10pt}
\noindent where the rows and columns of matrix $\mc{Q}^{ijkl}_{m,n}\equiv \mc{Q}^{ij,kl}_{m,n}$ are ordered such that $ij$ and $kl$ take the sequence of 11, 22, 12, 21, as noted in the equation above. This result will be used in the next subsection for the single $\mathbb{Z}_2$ theory. For the case of double $\mathbb{Z}_2$ theory, we have an additional symmetry of $\phi_1 \leftrightarrow \phi_2$, which implies ${\cal Q}_{(a)}^{1111}={\cal Q}_{(a)}^{2222}$. Therefore, a simple basis for $\mc{Q}^{ijkl}_{m,n}$ with the double $\mathbb{Z}_2$ symmetries $\phi_1 \leftrightarrow \phi_2$ and $\phi_1\to -\phi_1$ is given by
\be
\mc{Q}_{(1)}=\begin{pmatrix}
&1&0&0&0\\&
0&1&0&0\\&
0&0&0&0\\&
0&0&0&0
\end{pmatrix}
,~~~~~
\mc{Q}_{(2)}=\begin{pmatrix}
&0&1&0&0\\&
1&0&0&0\\&
0&0&0&1\\&
0&0&1&0
\end{pmatrix}
,~~~~~
\mc{Q}_{(3)}=\begin{pmatrix}
&0&0&0&0\\&
0&0&0&0\\&
0&0&1&0\\&
0&0&0&1
\end{pmatrix}
,~~~
\mc{Q}_{(-1)}=\begin{pmatrix}
&0&1&0&0\\&
1&0&0&0\\&
0&0&0&-1\\&
0&0&-1&0
\end{pmatrix}   .
\label{QdoubleZ2}
\ee
For $\mc{N}^{ijkl}_{(a)}$, the conditions $\mc{N}_{p,q}^{ijkl}=\mc{N}_{p,q}^{ilkj}=\mc{N}_{p,q}^{kjil}=\mc{N}_{m,n}^{jilk}$ and the double $\mathbb{Z}_2$ symmetries require that $\mc{N}^{ijkl}_{p,q}$ have the same symmetries as $\mc{Q}^{ijkl}_{2h,n}$, so we shall choose $\mc{N}^{ijkl}_{p,q}$ to have the same basis as $\mc{Q}^{ijkl}_{2h,n}$. This means that we will use the following null constraints
\bal
N^{(\mc{N})}_{p,q}&= z^1_{p,q}\[(2E^+_{p,q}+2F^+_{p,q})\mc{Q}_{(1)}\] +z^2_{p,q}\[(2E^+_{p,q}+F^+_{p,q})\mc{Q}_{(2)}+2F^+_{p,q}\mc{Q}_{(3)}-F^-_{p,q}\mc{Q}_{(-1)}\]
\nn
&~~~~ +z^3_{p,q}\[2E^+_{p,q}\mc{Q}_{(3)}+F^+_{p,q}\mc{Q}_{(2)}+F^-_{p,q}\mc{Q}_{(-1)}\]    .
\eal
Our empirical numerical explorations also confirm that including additional null constraints does not affect the positivity bounds. For the single scalar, we would only have the null constraint $\[(2E^+_{p,q}+2F^+_{p,q})\mc{Q}_{(1)}\]$ at any given $p$ and $q$, but for the double $\mathbb{Z}_2$ bi-scalar case we have two other null constraints at the same $p$ and $q$. (Note that this is not at any given $p+q$, which would also have multiple null constraints even for the single scalar case.) Since all the basis matrices are block diagonal, the linear matrix inequality of this SDP takes a simple block diagonal form
\be
\begin{pmatrix}
&A&0_{2\times2}\\&
0_{2\times2}&B
\end{pmatrix}\succeq 0    ,
\ee
where $A$ and $B$ are $2\times 2$ matrices. This is obviously equivalent to the lower dimensional conditions $A\succeq 0$ and $B\succeq 0$.

With the bases established, we can run the SDP (\ref{sdpPractical}) with {\nb SDPB} to decide whether a given set of Wilson coefficients in a particular problem can satisfy the triple crossing positivity bounds. To carve out the geometric shapes of the bounds for a given set of Wilson coefficients, we search for places where the objective function vanishes to find the boundaries of the positivity bounds.

\subsubsection*{$v^2$ order coefficients}

Let us first determine the finiteness of the $v^2$ coefficients. The optimal lower positivity bounds for the $v^2$ coefficients in a multi-field EFT can be identified as the extremal rays of the dual cone of the $v^2$ coefficient cone, and generally these optimal bounds can be obtained by normal SDPs with no continuous decision variable \cite{Li:2021lpe}. For double $\mathbb{Z}_2$ bi-scalar theory, these optimal bounds can also be obtained analytically and are described by the three inequalities \cite{Li:2021lpe}
\be
\label{v2cone}
{c}^{2,0}_{(1)} \geq 0, ~~ {c}^{2,0}_{(3)} \geq 0,~~ {c}^{2,0}_{(1)} \geq \pm  {c}^{2,0}_{(2)}-{c}^{2,0}_{(3)}   .
\ee
On the other hand, the 3D upper bounds on the ${c}^{2,0}_{(i)}$ coefficients of a double $\mathbb{Z}_2$ bi-scalar theory are computed in \cite{Hong:2024fbl} by formulating it as an SDP problem, as mentioned above around \eqref{upperSDP}. The results are obtained using null constraints of the lowest orders and can be further improved. Nevertheless, combining with the lower bounds and projected to 1D bounds, the two-sided bounds on the ${c}^{2,0}_{(i)}$ coefficients are given by
\bal
\label{adhoc}
0\leq {c}^{2,0}_{(1)}\leq 3.19(4\pi)^2,~~ - 1.32(4\pi)^2 \leq {c}^{2,0}_{(2)}\leq 2.46(4\pi)^2,~~0\leq {c}^{2,0}_{(3)}\leq 1.54(4\pi)^2   .
\eal

\subsubsection*{$v^2t$ order coefficients}

Now, we move away from the forward limit and consider coefficients with a $t$ derivative. In this subsection, for simplicity, we shall focus on the subspace where all the coefficients associated with $\mc{Q}_{(3)}$ vanish: $c_{(3)}^{m,n}=0$, which allows us to visualize the bounds. This means that we still have $\mc{Q}_{(3)}$ and associated $y$ in the SDP, but we simply set $c_{(3)}^{m,n}=0$ in the objective function. 

Generally, there are two independent Wilson coefficients for any given $n$ and even $m=2h$ in the $c_{(3)}^{m,n}=0$ subspace of the double $\mathbb{Z}_2$ theory: $c_{(1)}^{2h,n}=2c_{1111}^{m,n},~c_{(2)}^{2h,n}=4c_{1122}^{2h,n}$. So we shall investigate the positivity bounds on the coefficients:  $c_{(1)}^{2,0}$, $c_{(2)}^{2,0}$, $c_{(3)}^{2,0}$, $c_{(1)}^{2,1}$ and $c_{(2)}^{2,1}$. For this SDP problem, the components of the $A$ and $B$ matrices are explicitly given by
\bal
\label{dz2A}
&A_{11}=A_{22}=\displaystyle y^1_{2,0} \f{C_\ell^{2,0}}{ \mu^{3}}+y^1_{2,1}\f{C_\ell^{2,1}}{\mu^{4}}+\sum_{p,q} \f1{\mu^{p+q+1}}[z^1_{p,q}(2E^+_{p,q}+2F^+_{p,q})]   \\&
A_{12}=A_{21}=\displaystyle y^2_{2,0}\f{C_\ell^{2,0}}{\mu^{3}}+y^2_{2,1}\f{C_\ell^{2,1}}{\mu^{4}}+\sum_{p,q} \f1{\mu^{p+q+1}}[z^2_{p,q}(2E^+_{p,q}+F^+_{p,q}-F^-_{p,q})+z^3_{p,q}(F^+_{p,q}+F^-_{p,q})] \\&
\label{dz2B}
B_{11}=B_{22}= y^3_{2,0}\f{C_\ell^{2,0}}{\mu^{3}}+y^3_{2,1} \f{C_\ell^{2,1}}{\mu^{4}}+\sum_{p,q} \f1{\mu^{p+q+1}}[z^2_{p,q}(2F^+_{p,q})+z^3_{p,q}(2E^+_{p,q})]\\&
B_{12}=B_{21}= y^2_{2,0}\f{C_\ell^{2,0}}{\mu^{3}}+y^2_{2,1} \f{C_\ell^{2,1}}{\mu^{4}}+\sum_{p,q} \f1{\mu^{p+q+1}}[z^2_{p,q}(2E^+_{p,q}+F^+_{p,q}+F^-_{p,q})+z^3_{p,q}(F^+_{p,q}-F^-_{p,q})]   .
\eal  
As mentioned, we have kept $y^3_{2,0}$ and $y^3_{2,1}$ in the $A$ and $B$ matrices, but set $c_{(3)}^{2,0}=c_{(3)}^{2,1}=0$ in the objective function --- we are not agnostic on the value of $c_{(3)}^{2,0}$ and $c_{(3)}^{2,1}$ here. We can normalize all $c_{(a)}^{m,n}$ coefficients with $c_{(1)}^{2,0}$, defining the coefficients with a tilde
\be
\td c_{(a)}^{m,n} \equiv \f{ c_{(a)}^{m,n} }{c_{(1)}^{2,0}} ,
\ee 
which leaves us with 3 parameters $(\td c_{(2)}^{2,0},~\td c_{(1)}^{2,1},~\td c_{(2)}^{2,1})$. Additionally, the choice of looking at the $c_{(3)}^{2,0}=c_{(3)}^{2,1}=0$ subspace allows us to ignore the $B\succeq 0$ condition. This is because, if $A\succeq 0$ is feasible, we see from the explicit expressions above that the two conditions of $B\succeq 0$ ($B_{11}=B_{22}\geq 0$ and ${\rm det}B=B_{11}^2-B_{12}^2\geq 0$) do not affect the SDP results, as they can always be solved by appropriate $y^3_{2,0}$ and $y^3_{2,1}$ ($y^3_{2,0}$ and $y^3_{2,1}$ are linearly bundled together with $y^2_{2,0}$ and $y^2_{2,0}$ respectively), which does not affect the objective function where $c_{(3)}^{2,0}=c_{(3)}^{2,1}=0$. Therefore, we only need to solve the $A\succeq 0$ condition.

\begin{figure}[h]
\bc
\includegraphics[width=0.55\textwidth]{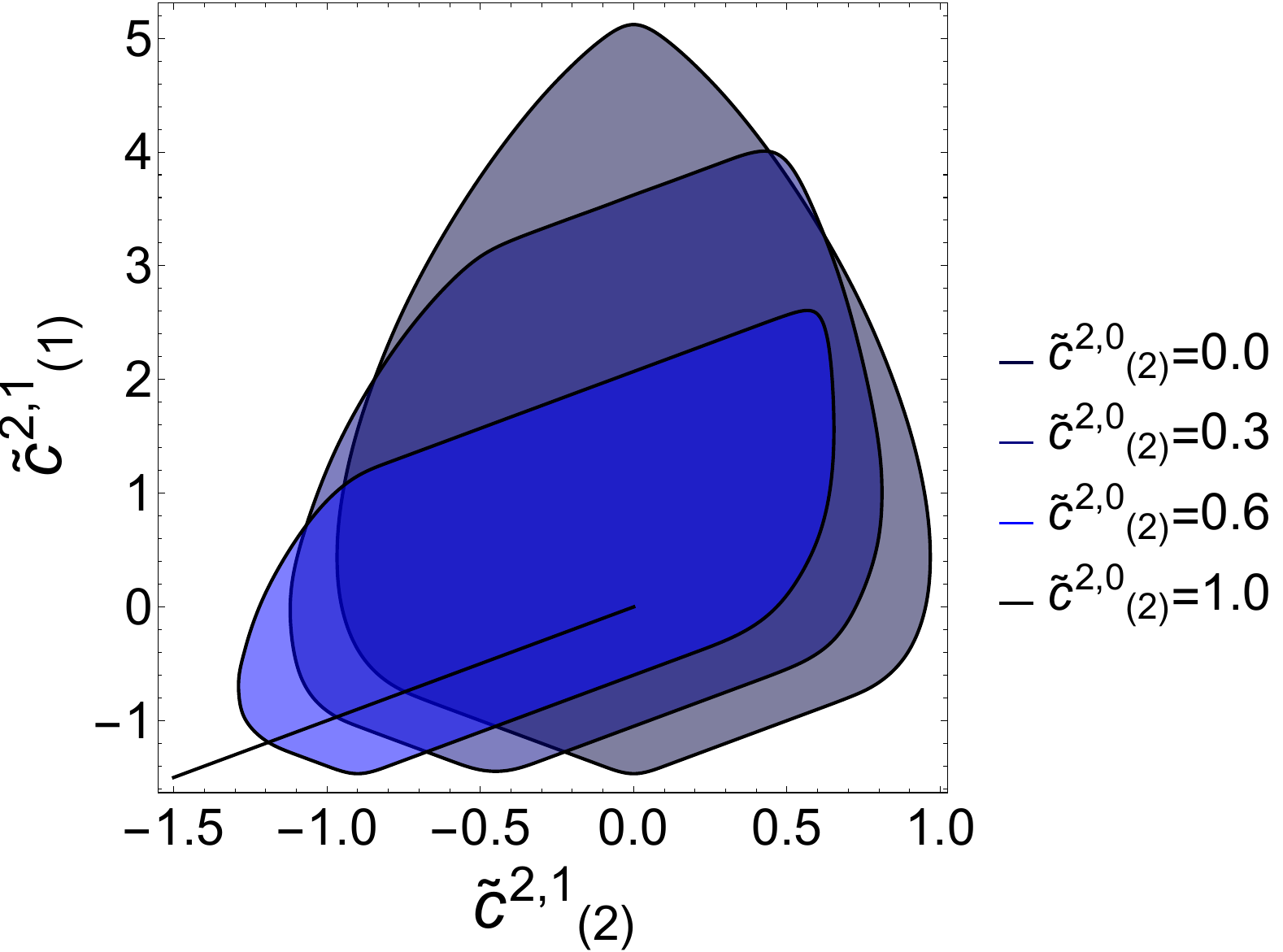}
\caption{Positivity bounds on $(\td c_{(2)}^{2,0}=4\td c^{2,0}_{1122},\td c_{(1)}^{2,1}=2\td c_{1111}^{2,1},\td c_{(2)}^{2,1}=4\td c^{2,1}_{1122})$ for bi-scalar theory with double $\mathbb{Z}_2$ symmetry in the $\td c^{m,n}_{(3)}=2\td c^{m,n}_{1221}=0$ subspace, agnostic about all higher order coefficients. $c_{ijkl}^{m,n}$ is the Wilson coefficient in front of the $(s+\f{t}2)^m t^n$ term in the $ij\to kl$ amplitude, and the tilded coefficients are defined as $\td c_{(a)}^{m,n}\equiv c_{(a)}^{m,n}/c_{(1)}^{2,0}$. We choose units such that the cutoff $\Lambda=1$. We see that the triple crossing positivity bounds form a closed, finite region in the parameter space, numerically of order $\mc{O}(1)$.}
\label{fig:doubleZ2even}
\ec
\end{figure}

Before presenting our numerical evaluations of the 3D positivity bounds, notice that, from the result of the $v^2$ convex cone (\ref{v2cone}), we know that $\td c^{2,0}_{(2)}$ must stay in the range $-1\leq \td c^{2,0}_{(2)}\leq 1$ in the subspace $\td c_{(3)}^{2,0}=0$. Also, the SDP problem above is symmetric under $\td c^{2,0}_{(2)} \to -\td c^{2,0}_{(2)}$ and $\td c^{2,1}_{(2)} \to -\td c^{2,1}_{(2)}$. So it is sufficient to sample from $0$ to $1$ for the $\td c^{2,0}_{(2)}$ parameter. See Figure \ref{fig:doubleZ2even} for a few slices of this 3D space, which deforms from an ``inverted heart'' shape to a line segment from  $\td c^{2,0}_{(2)}=0.0$ to $\td c^{2,0}_{(2)}=1.0$. We can see that, similar to the single scalar case, the Wilson coefficients are constrained in a small finite region, parametrically around $\mc{O}(1)$ in the units of $c^{2,0}_{(1)}$.

\subsubsection*{Higher order coefficients}

When moving away from the $c_{(3)}=0$ subspace, the triple crossing bounds are still enclosed, as we shall see now. However, to see that, we need to additionally make use of the upper bounds of the $v^2$ coefficients. In this subsection, we shall compute two-sided bounds for the $c_{(a)}^{m\geq 2,n>0}$ coefficients. The two-sided bounds for $c_{(1)}^{m\geq 2,n>0}$ are basically the ones for the single scalar case, for which we can also compute two-sided bounds for $\td c_{(1)}^{m\geq 2,n>0}$, the ratio between $c_{(1)}^{m\geq 2,n>0}$ and $c_{(1)}^{2,0}$. However, we find that the ratio between $c_{(a\neq 1)}^{m\geq 2,n>0}$ and $c_{(1)}^{2,0}$ is generally unbounded. Nevertheless, as we shall see, $c_{(a\neq 1)}^{m\geq 2,n>0}$ itself is indeed bounded from both sides. This is not surprising as, for the multi-field case, the positivity bounds on the $v^2$ coefficients form a multi-dimensional shape and the ratios between themselves are not even fully bounded. Apart from $c_{(2)}^{2h,n}$ and $c_{(3)}^{2h,n}$, we also have another coefficient $c_{(-1)}^{2h+1,n}=4c_{1122}^{2h+1,n}$ for any given $n$ and odd $m=2h+1$, so in this subsection we want to compute the two-sided bounds for the $c_{(2)}^{2h,n}$, $c_{(3)}^{2h,n}$ and $c_{(-1)}^{2h+1,n}$ coefficients. 

\begin{figure}[h]
\bc
\includegraphics[width=0.4\textwidth]{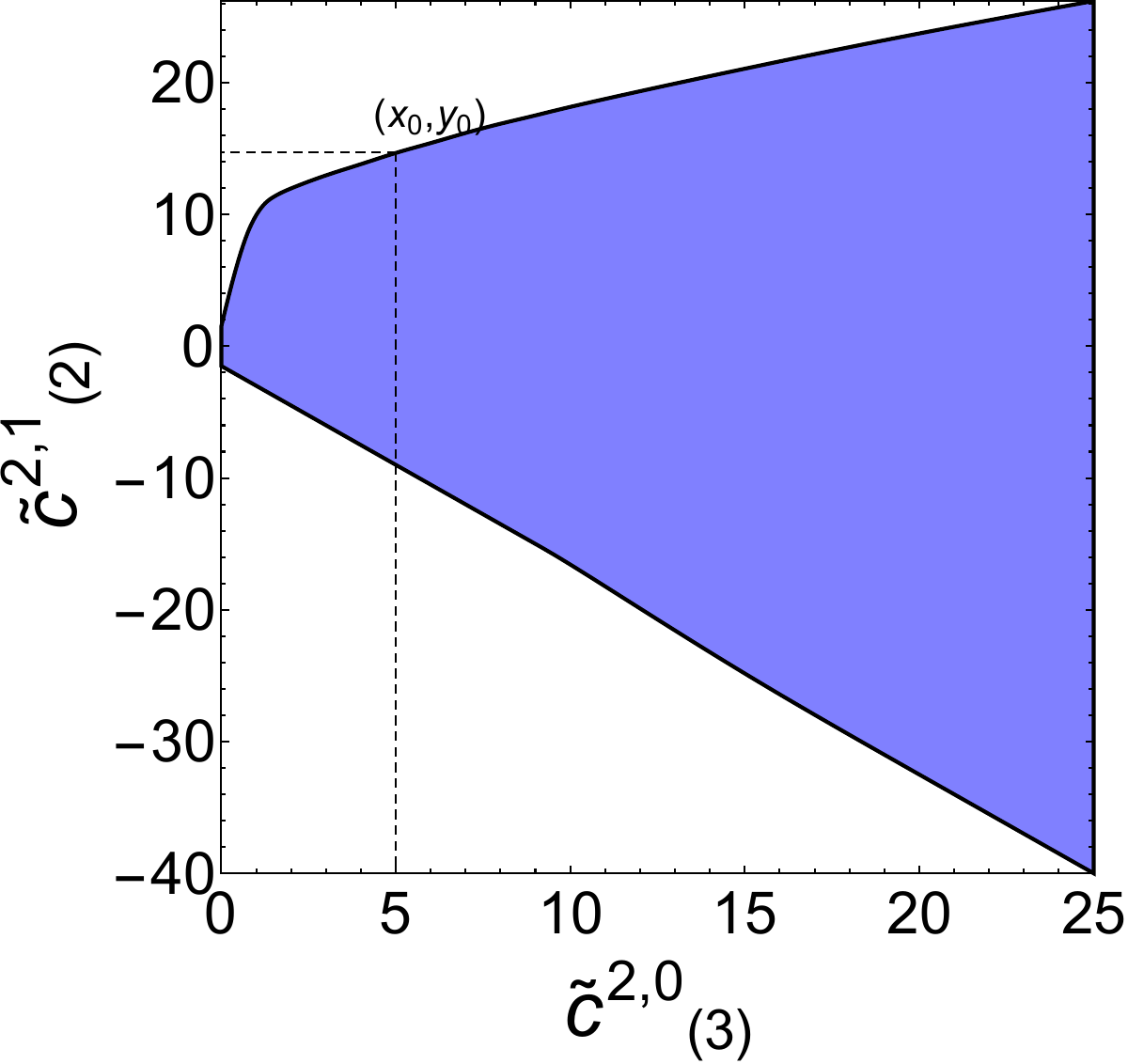}
~~~~~~~
\includegraphics[width=0.4\textwidth]{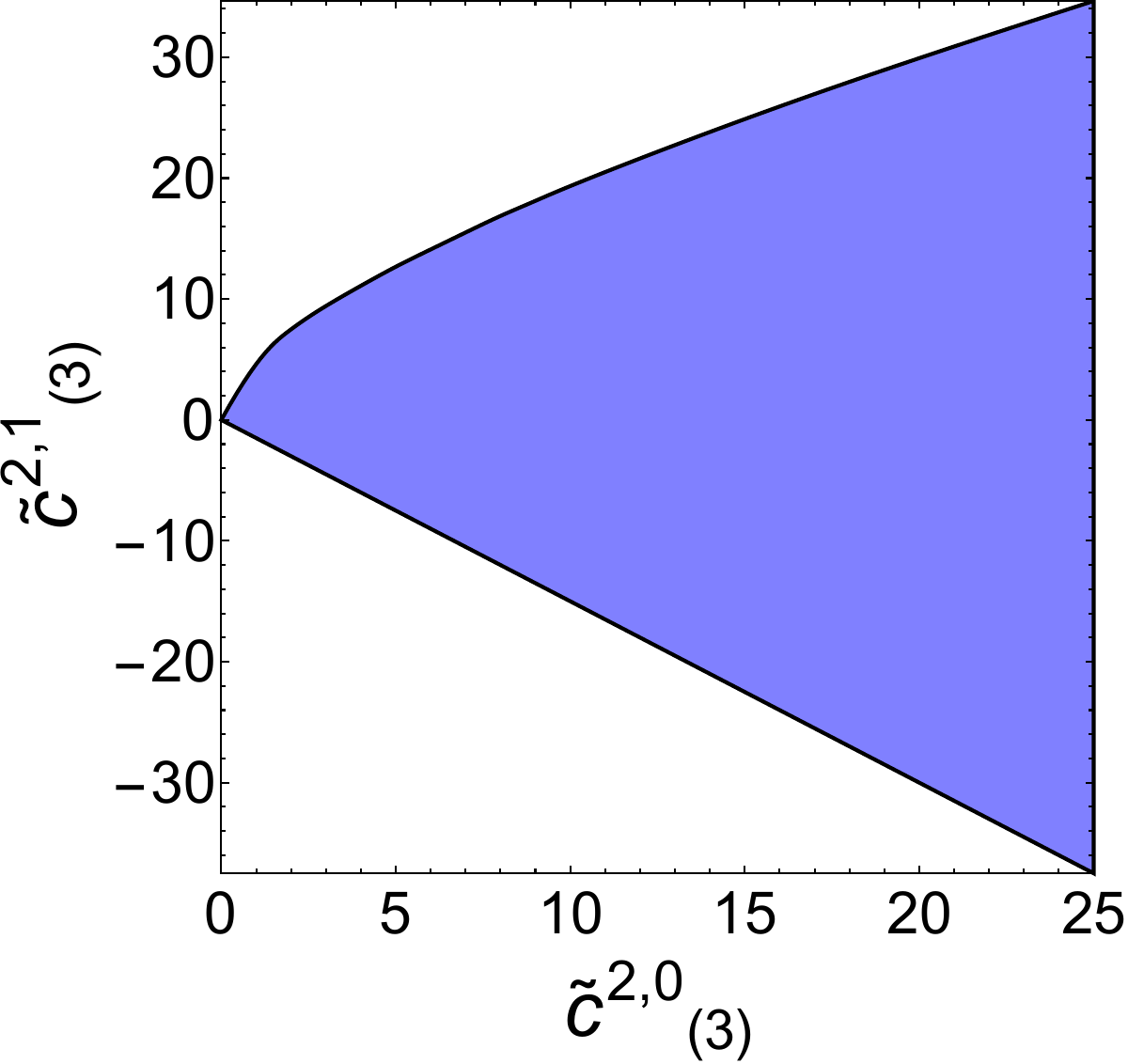}
\caption{Positivity bounds on $(\tilde{c}_{(3)}^{2,0}=2\tilde{c}_{1212}^{2,0},\tilde{c}_{(2)}^{2,1}=4\tilde{c}^{2,1}_{1122})$ and $(\tilde{c}_{(3)}^{2,0}=2\tilde{c}_{1212}^{2,0},\tilde{c}_{(3)}^{2,1}=2\tilde{c}^{2,1}_{1212})$ for double $\mathbb{Z}_2$ symmetric bi-scalar theory. The SDP includes 3 parameters, $c_{(1)}^{2,0}$, $c_{(3)}^{2,0}$ and $c_{(2)}^{2h,n}$, and is agnostic about all other coefficients. In both the left and right plot, the right hand side of the bound extends to infinity.}
\label{fig:z2c2c3}
\ec
\end{figure}

Let us start with $c_{(2)}^{2,1}$. If we solve SDPs for $\td c_{(2)}^{2,1}=c_{(2)}^{2,1}/c_{(1)}^{2,0}$, agnostic about all other coefficients, we find it unbounded. (The same applies if we look for bounds on the ratio $c_{(2)}^{2,1}/c_{(2)}^{2,0}$ or $c_{(2)}^{2,1}/c_{(3)}^{2,0}$.) To see this more clearly, let us draw a 2D positivity bound region for $\td c^{2,1}_{(2)}$ and $\td c^{2,0}_{(3)}$; see the left plot of Figure \ref{fig:z2c2c3}. (A similar plot can be drawn for $\td c_{(3)}^{2,1}$ and $\td c_{(3)}^{2,0}$; see the right plot of Figure \ref{fig:z2c2c3}.) As we can see in this plot, if we look for the triple crossing bounds on $\td c_{(2)}^{2,1}=c_{(2)}^{2,1}/c_{(1)}^{2,0}$, agnostic about all other coefficients, it is simply unbounded from below and above, which remains true even if we include the upper bound for $c_{(1)}^{2,0}$. This is because, although the $v^2$ coefficients are constrained in a finite region, the ratio $\td c_{(3)}^{2,0} = c_{(3)}^{2,0}/c_{(1)}^{2,0}$ is unbounded from above, which is easy to understand as the positivity region contains points where $c_{(1)}^{2,0}$ approaches to zero but $c_{(3)}^{2,0}$ remains finite. Nevertheless, we can look for the bounds on $c_{(2)}^{2,1}$ itself, and once we also include the generalized upper bounds (\ref{adhoc}), we can constrain $c_{(2)}^{2,1}$ from both sides. To see how this works, let us look at the example of the upper boundary of the left plot of Figure \ref{fig:z2c2c3}, which starts from near the origin, intersecting with the positive $\td c^{2,1}_{(2)}$ axis, and extends north-eastward to infinity. Suppose $(x_0, y_0)$ is an arbitrary point on this boundary line. For all the points in the positivity region where $\td c^{2,0}_{(3)}< x_0$, we have  
\be
\td c^{2,1}_{(2)} = \f{ c^{2,1}_{(2)}}{ c^{2,0}_{(1)}} <y_0    .
\ee
Since $0<c^{2,0}_{(1)}< 3.19 (4\pi)^2$, we can infer that $c^{2,1}_{(2)}<y_0c^{2,0}_{(1)}< 3.19(4\pi)^2y_0$. For all the points in the positivity region where $\td c^{2,0}_{(3)}> x_0$, on the other hand, we have
\be
\f{c^{2,1}_{(2)}}{ c^{2,0}_{(3)}}=\f{ \td c^{2,1}_{(2)}}{ \td c^{2,0}_{(3)}} < \f{y_0}{x_0}    ,
\ee
thanks to the fact that the boundaries of positivity bounds are convex. Since $0<c^{2,0}_{(3)}<1.54(4\pi)^2$, we can infer that $c^{2,1}_{(2)}<y_0c^{2,0}_{(3)}/x_0< 1.54(4\pi)^2y_0/x_0$. So $c^{2,1}_{(2)}$ must be bounded by the maximum of $3.19(4\pi)^2y_0$ and $1.54(4\pi)^2 {y_0}/{x_0}$. The optimal upper bound for $c^{2,1}_{(2)}$ is obtained by searching over all $x_0$ so as to get the smallest upper bound. Since the upper boundary is convex and monotonically increasing with $x_0$, we can infer that $3.19(4\pi)^2y_0$ increases and $1.54(4\pi)^2 {y_0}/{x_0}$ decreases as $x_0$ increases, so the optimal bound is reached at $3.19 y_0=1.54{y_0}/{x_0}$, \ie  $\text{min}_{x_0}( \text{max} (y_0,0.48{y_0}/{x_0}) )=y_0|_{x_0=0.48}$. Therefore, we have the optimal upper bound
\be
c^{2,1}_{(2)} < {3.19}(4\pi)^2 y_0|_{x_0=0.48}    .
\ee
Similarly, we can find a lower bound for $c^{2,1}_{(2)}$ by considering the lower boundary in the left plot of Figure \ref{fig:z2c2c3}, which is given by $c^{2,1}_{(2)} > 3.19(4\pi)^2 y_0|_{x_0=0.48}$.

We can apply the above procedure to obtain the two-sided bounds for all the $c_{(2)}^{2h,n}$, $c_{(3)}^{2h,n}$ and $c_{(-1)}^{2h+1,n}$ coefficients, in units of appropriate powers of the cutoff $\Lambda$. Note that for all these coefficients, in the plot similar to those in Figure \ref{fig:z2c2c3}, the upper (lower) boundary must intersect with the positive (negative) axis at $\tilde c_{(3)}^{2,0}=0$, as these coefficients being zero must be within the positivity bounds. The two-sided bounds for the first few coefficients in double $\mathbb{Z}_2$ bi-scalar theory can be found in Table \ref{tab:1} below:
\be
\label{LUdef}
 L^{(2)}_{2h,n} < \f{c_{(2)}^{2h,n} }{(4\pi)^2} < U^{(2)}_{2h,n},~~~~~~  L^{(3)}_{2h,n} < \f{c_{(3)}^{2h,n}}{(4\pi)^2}  < U^{(3)}_{2h,n},~~~~~~  L^{(-1)}_{2h+1,n} < \f{c_{(-1)}^{2h+1,n}}{(4\pi)^2}  < U^{(-1)}_{2h+1,n}   ,
\ee
which are obtained by truncating the null constraints below $p+q=9$ and the partial wave spin below $\ell=50$ but additionally including the $\ell =\infty$ partial wave to speed up the convergence.

\begin{table}[h!]
\centering
{\footnotesize
\begin{equation*}
\begin{array}{|c|c|c|c|c|c|c|c|c|c|c|c|}
\hline (2h,n) & (2,1) & (2,2) & (4,0) & (2,3) & (4,1) & (2,4) & (4,2) & (6,0) & (2,5) & (4,3) & (6,1) \\

\hline
L^{(2)}_{2h,n} & -7.08  & -8.36  & -4.72 & -6.67  & -11.8 & -2.99  & -21.5 & -4.72 & -9.54  & -20.6  & -16.5  \\
\hline U^{(2)}_{2h,n} &22.3 &7.08  &4.72 &8.39  &18.2 &17.1 &17.7 &4.72 &4.63  &19.2  &15.3  \\
\hline

\hline L^{(3)}_{2h,n} & -2.30 & -2.30 & 0.00 & -2.19 & -3.83 & -2.47 & -7.69 & 0.000 & -3.38 & -8.64 &-5.36  \\
\hline U^{(3)}_{2h,n} & 10.8 & 14.4 & 1.53 & 11.6 & 8.93  & 11.9  & 9.51 & 1.53 & 11.2  & 6.89  & 7.40  \\
\hline
\hline (2h+1,n)  & (3,1) & (3,2) & (5,0) & (3,3) & (5,1) & (3,4) & (5,2) & (7,0) &  &  &  \\
\hline

\hline L^{(-1)}_{2h+1,n} & -6.38   & -14.0 & -4.72  & -7.98   & -11.1   & -12.8  & -26.8 & -4.72  &  &  &  \\
\hline U^{(-1)}_{2h+1,n} & 23.5 & 8.07 & 4.72 & 12.4 & 20.5 & 25.0 & 16.7 & 4.72 &  &  &  \\
\hline
\end{array}
\end{equation*} 
}
\vskip -15pt
\caption{Two-sided bounds for $c_{(2)}^{2h,n}$, $c_{(3)}^{2h,n}$ and $c_{(-1)}^{2h+1,n}$, for which we are agnostic about all the other coefficients except for the $v^2$ coefficients. $L^{(a)}_{m,n}$ and $U^{(a)}_{m,n}$ are defined in \eref{LUdef}. We use the units where the cutoff $\Lambda=1$.}
\label{tab:1}
\end{table}

\subsection{Bi-scalar theory with $Z_2$ symmetry}

In the last subsection, we discussed one of the simplest examples of multi-field full crossing bounds and found that the positivity bounds constrain the Wilson coefficients to a small finite region close to the origin. In this subsection, we will slightly increase the complexity by relaxing the symmetries of the theory to consider a bi-scalar theory with $Z_2$ symmetry
\be
\phi_i\rightarrow -\phi_i,~~i=1,2   .
\ee
We will see that the Wilson coefficients can still be constrained to a small finite region. 

For the $Z_2$ bi-scalar theory, $\mc{Q}^{ijkl}_{m,n}$ must take the form of \eref{Qmnform}, and so its basis can be chosen as 
\bal
\mc{Q}_{(1)} =\begin{pmatrix}
&1&0&0&0\\&
0&0&0&0\\&
0&0&0&0\\&
0&0&0&0
\end{pmatrix}
,~~~
\mc{Q}_{(2)} = &\begin{pmatrix}
&0&0&0&0\\&
0&1&0&0\\&
0&0&0&0\\&
0&0&0&0
\end{pmatrix}
,~~~
\mc{Q}_{(3)}=\begin{pmatrix}
&0&1&0&0\\&
1&0&0&0\\&
0&0&0&1\\&
0&0&1&0
\end{pmatrix}
\\
\mc{Q}_{(4)}=\begin{pmatrix}
&0&0&0&0\\&
0&0&0&0\\&
0&0&1&0\\&
0&0&0&1
\end{pmatrix}
&,~~~
\mc{Q}_{(-1)}=\begin{pmatrix}
&0&1&0&0\\&
1&0&0&0\\&
0&0&0&-1\\&
0&0&-1&0
\end{pmatrix}   .
\eal
With this basis, we have $c_{(1)}^{m, n}=c_{1111}^{m, n}, ~ c_{(2)}^{m, n}=c_{2222}^{m, n}, ~ c_{(3)}^{m, n}=4 c_{1122}^{m, n}, ~ c_{(4)}^{m, n}=2 c_{1212}^{m, n}$, and other Wilson coefficients are related to the above ones by crossing and internal symmetries. Again, we choose $\mc{N}^{ijkl}_{p,q}$ to have the same basis as $\mc{Q}^{ijkl}_{2h,n}$. From these bases, we see that the linear matrix inequality can still be cast in a block diagonal form
\be
\left(\begin{array}{cc}
A & 0_{2 \times 2} \\
0_{2 \times 2} & B
\end{array}\right) \succeq 0   .
\ee

First, we shall explore the shape of the full crossing symmetric bounds on the coefficients of the $v^2$ and $v^2t$ terms in the $\mathbb{Z}_2$ theory. Truncated to this next leading order, we already have 7 parameters: $\td c_{(2)}^{2,0}$, $\td c_{(3)}^{2,0}$, $\td c_{(4)}^{2,0}$, $\td c_{(1)}^{2,1}$, $\td c_{(2)}^{2,1}$, $\td c_{(3)}^{2,1}$ and $\td c_{(4)}^{2,1}$, if we measure them in terms of $c_{(1)}^{2,0}$. For simplicity, we shall restrict to the subspace $\td c_{(4)}^{2,0}=\td c_{(4)}^{2,1}=0$, for which case, the $B\succeq 0$ condition again can be neglected. Then, the SDP problem additionally contains a scaling invariance in the remaining 5D parameter space: $c_{(2)}\rightarrow \ai^2c_{(2)},~c_{(3)}\rightarrow \ai c_{(3)}$. To see this, note that the $A$ matrix is explicitly given by
\be
A=\begin{pmatrix}
&y^1_{2,0} \f{C_\ell^{2,0}}{ \mu^{3}}+y^1_{2,1}\f{C_\ell^{2,1}}{\mu^{4}}+ (\text{null-1})  
&y^3_{2,0}\f{C_\ell^{2,0}}{\mu^{3}}+y^3_{2,1}\f{C_\ell^{2,1}}{\mu^{4}}+ (\text{null-2})
\\
& y^3_{2,0}\f{C_\ell^{2,0}}{\mu^{3}}+y^3_{2,1} \f{C_\ell^{2,1}}{\mu^{4}}+ (\text{null-2})
&y^2_{2,0} \f{C_\ell^{2,0}}{\mu^{3}}+y^2_{2,1}\f{C_\ell^{2,1}}{\mu^{4}}+  (\text{null-3})
\end{pmatrix}   ,
\ee
where $(\text{null-1})$ to $(\text{null-4})$ are four different sums of the null constraints. $A$ being semi-definite positive is equivalent to $A_{11} \geq 0$, $A_{22} \geq 0$ and $A_{11} A_{22} \geq (A_{12})^2$, and the objective is to minimize $y^1_{2,0}c_{(1)}^{2,0}+y^2_{2,0}c_{(2)}^{2,0}+y^3_{2,0}c_{(3)}^{2,0}+y^2_{2,1}c_{(2)}^{2,1}+y^3_{2,1}c_{(3)}^{2,1}$. Clearly, we can re-write the objective and the constraints as 
\bal
\text{minimize }&y^1_{2,0}+\bar y^2_{2,0} \(\ai^2c_{(2)}^{2,0}\)+\bar y^3_{2,0}\(\ai c_{(3)}^{2,0}\)+\bar y^2_{2,1}\(\ai^2 c_{(2)}^{2,1}\)+\bar y^3_{2,1}\(\ai c_{(3)}^{2,1}\)
\\
\text{subject to }&A_{11} \geq 0,~~\( A_{22}|_{y^2\to \bar y^2, z^2\to \bar z^2}\) \geq 0,~~A_{11} \( A_{22}|_{y^2\to \bar y^2, z^2\to \bar z^2}\) \geq \(A_{12}|_{y^3\to \bar y^3, z^3\to \bar z^3}\)^2    ,
\eal
where we have defined $\bar y^2\equiv y^2/\ai^2, ~\bar y^3\equiv y^3/\ai$, and $|_{y^2\to \bar y^2}$, for example, means replacing $y^2_{2,0}$ and $y^2_{2,1}$ in $A_{22}$ with $\bar y^2_{2,0}$ and $\bar y^2_{2,1}$ respectively. The scaled SDP is the same as the original SDP but with scaled coefficients $\ai^2c_{(2)}$ and $\ai c_{(3)}$. We can use this scaling invariance to set $\td c_{(2)}^{2,0}=1$ without loss of generality, and we are left 4 Wilson coefficients $\td c_{(3)}^{2,0}$, $\td c_{(1)}^{2,1}$, $\td c_{(2)}^{2,1}$, $\td c_{(3)}^{2,1}$.

Of course, we still can not visualize 4D parameter space, so we will plot the positivity bounds for different $\td c_{(3)}^{2,0}$. From the results of the $v^2$ convex cone, we know that viable $\td c_{(3)}^{2,0}$ must be in the range \cite{Li:2021lpe}: $-2\leq \td c_{(3)}^{2,0} \leq 2$, which is symmetric with respect to $\td c_{(3)}^{2,0}=0$. To probe how the positivity region varies with different $\td c_{(3)}^{2,0}$, we shall look at 3 slices of the 4D parameter space, and we will see that on all of them the 3D parameter space $(\td c_{(1)}^{2,1},~\td c_{(2)}^{2,1},~\td c_{(3)}^{2,1})$ is constrained to a finite region.

\begin{itemize}

\item {\it The $\td c_{(3)}^{2,0}=2.0$ slice:} Since it is at one of the extremities of $\td c_{(3)}^{2,0}$, the allowed region is a line segment
\be
\td c^{2,1}_{(1)}=\td c^{2,1}_{(2)}=\f12 \td c^{2,1}_{(3)}=\eta,~~-1.5\leq\eta\leq 0   .
\ee

\begin{figure}[h]
\bc
\includegraphics[width=0.6\textwidth]{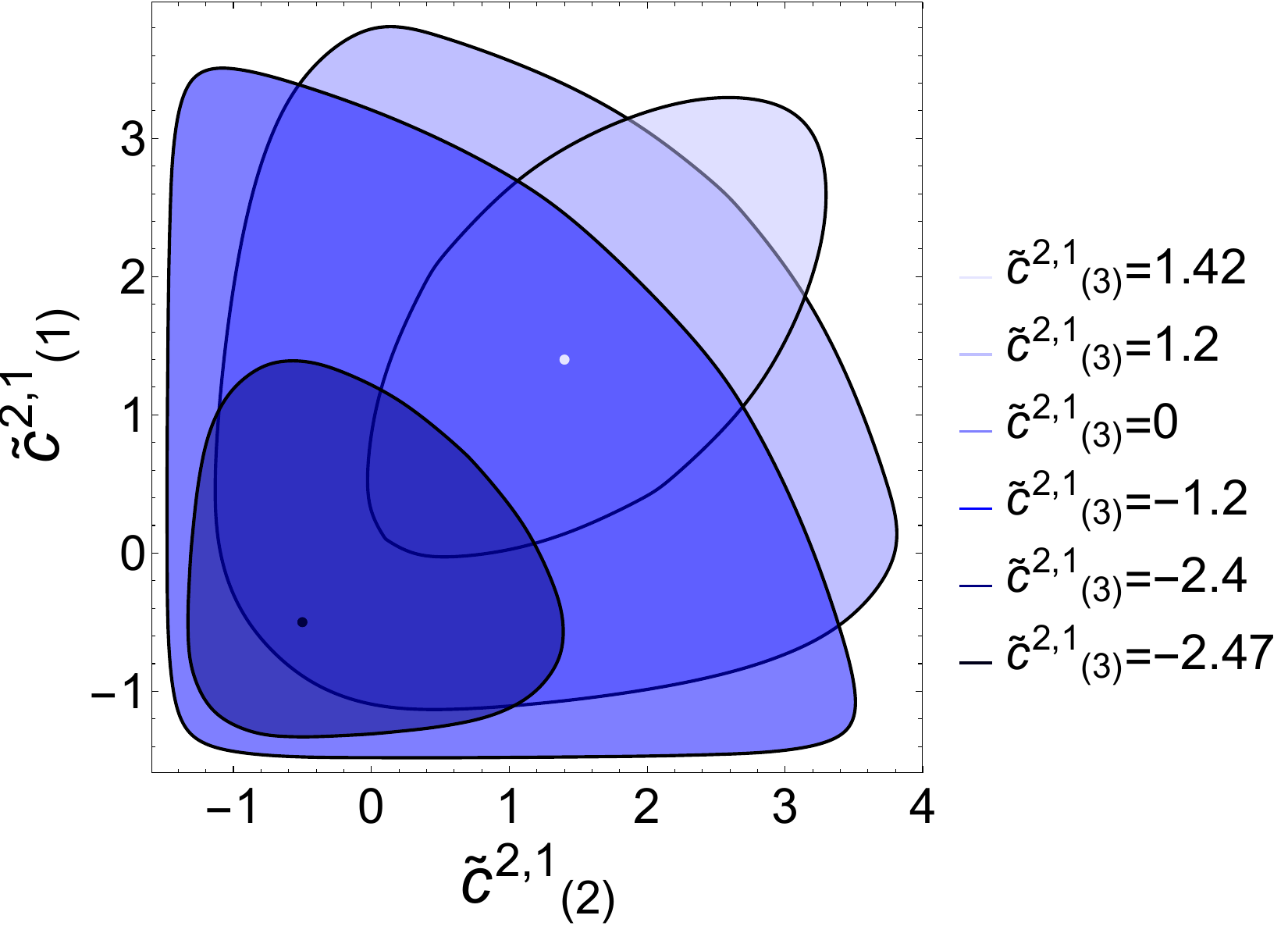}
\caption{Positivity bounds on $(\td c_{(1)}^{2,1}=\td c^{2,1}_{1111},\td c_{(2)}^{2,1}=\td c^{2,1}_{2222},\td c_{(3)}^{2,1}=4\td c^{2,1}_{1122})$ for $\mathbb{Z}_2$ symmetric bi-scalar theory on the $\td c_{(3)}^{2,0}=1.0$ slice. The tildes on the coefficients mean that they are divided by $c_{(1)}^{2,0}$. We have truncated up to order $(m=2, n=1)$, restricted to $\td c_{(4)}^{2,0}=\td c_{(4)}^{2,1}=0$ and set $\td c_{(2)}^{2,0}$ to 1 (without loss of generality). We choose units such that the cutoff $\Lambda=1$.}
\label{fig:z2full1}
\ec
\end{figure}

\item {\it The $\td c_{(3)}^{2,0}=1.0$ slice:} We now need to numerically running the SDP for many sets of Wilson coefficients to get contour plots for different values of $\td c_{(3)}^{2,1}$. Notice that the plots are symmetric in exchanging $\td c_{(1)}^{2,1}\leftrightarrow  \td c_{(2)}^{2,1}$, which is not surprising since we have chosen $\td c_{(2)}^{2,0}=1=\td c_{(1)}^{2,0}$. As shown in Figure \ref{fig:z2full1}, $\td c_{(3)}^{2,1}$ has an upper bound $\td c_{(3)}^{2,1}=1.42$ and a lower bound $\td c_{(3)}^{2,1}=-2.47$.

\begin{figure}[h]
\bc
\includegraphics[width=0.6\textwidth]{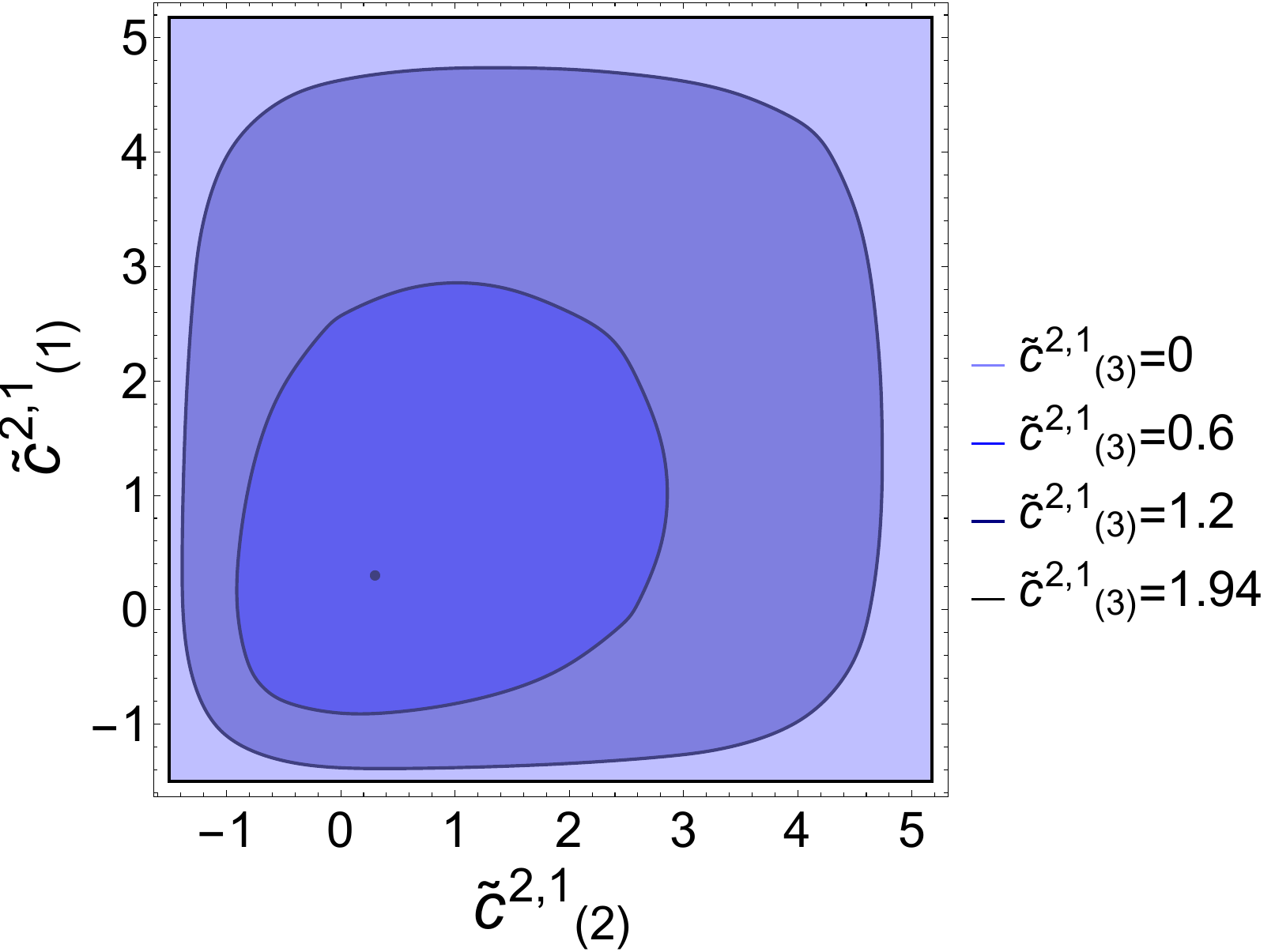}
\caption{Positivity bounds on $(\td c_{(1)}^{2,1}=\td c^{2,1}_{1111},\td c_{(2)}^{2,1}=\td c^{2,1}_{2222},\td c_{(3)}^{2,1}=4\td c^{2,1}_{1122})$ for $\mathbb{Z}_2$ symmetric bi-scalar theory on the $\td c_{(3)}^{2,0}=0.0$ slice.}
\label{fig:z2full2}
\ec
\end{figure}

\item {\it The $\td c_{(3)}^{2,0}=0.0$ slice:} We find that $\td c_{(1)}^{2,1}$ and $\td c_{(2)}^{2,1}$ reach both their lower limit ($-1.5$) and upper limit ($5.173$) when $\td c_{(3)}^{2,1}=0$, which are the box bounds in Figure \ref{fig:z2full2}, the 2D generalization of the two-sided bounds for single scalar scattering.  On the other hand, $\td c_{(3)}^{2,1}$ reaches both of its lower and upper limit at $\td c_{(3)}^{2,1}=\pm1.94$.
 
\end{itemize}

We can also consider two-sided bounds for the $\mathbb{Z}_2$ theory, \ie the bounds on $c^{2h,n}_{(3)}$, $c^{2h,n}_{(4)}$ and $c^{2h+1,n}_{(-1)}$. The two-sided bounds on $c^{2h,n}_{(1)}$ and $c^{2h,n}_{(2)}$ are the bounds in the single scalar case. Since the positivity region is convex, the positivity bounds must be symmetric in exchanging $c_{(1)}^{m,n}\leftrightarrow c_{(2)}^{m,n}$, and so the weakest bounds must occur in the subspace $c_{(1)}^{m,n}=c_{(2)}^{m,n}$. If we have $c_{(1)}^{m,n}=c_{(2)}^{m,n}$, we effectively reduce back to the double $\mathbb{Z}_2$ case. So the two-sided bounds on $c^{2h,n}_{(3)}$, $c^{2h,n}_{(4)}$ and $c^{2h+1,n}_{(-1)}$ here are exactly the same as those on $c^{2h,n}_{(2)}$, $c^{2h,n}_{(3)}$ and $c^{2h+1,n}_{(-1)}$ in the double $\mathbb{Z}_2$ theory. Notice that in our notation $c^{2h,n}_{(3)}$ and $c^{2h,n}_{(4)}$ here are the same as the $c^{2h,n}_{(2)}$ and $c^{2h,n}_{(3)}$ in the double $\mathbb{Z}_2$ theory.

\acknowledgments

We would like to thank Yu-tin Huang, David Simmons-Duffin, Ning Su and Zi-Yue Wang for helpful discussions. SYZ acknowledges support from the starting grants from University of Science and Technology of China under grant No.~KY2030000089 and GG2030040375, and is also supported by National Natural Science Foundation of China under grant No.~11947301, 12075233 and 12047502, and supported by the Fundamental Research Funds for the Central Universities under grant No.~WK2030000036. {\it We are very sad that Cen Zhang passed away in the middle of this work.} \\

\appendix

\section{Expressions for $\bar n^{p,q}_{ijkl}$, $E^{\pm}_{p,q}$ and $F^{\pm}_{p,q}$}
\label{sec:EF}

While the $jl$ symmetrized null constraints $n^{p,q}_{ijkl}=0$ are sufficient for our approach with the $su$ symmetric dispersive relation, it is conceivable that the unprojected null constraints $\bar n^{p,q}_{ijkl}=0$ can be useful in other approaches. Here we also list the first few $\bar n^{p,q}_{ijkl}$ null constraints for a comparison:
\bal
\label{nConstraintsStartfull}
\bar n_{ijkl}^{1,3}&=\bar n_{ijkl}^{2,2} = c_{ijkl}^{2,2}+\f32 (c_{ijkl}^{3,1}+ c_{ijkl}^{4,0})- c_{ikjl}^{2,2}-\f32(c_{ikjl}^{3,1}+c_{ikjl}^{4,0})=0
\\
\bar n_{ijkl}^{1,4}&=c_{ijkl}^{2,3}+\f34c_{ijkl}^{3,2}-\f5{8}c_{ijkl}^{5,0}+\f34c_{ikjl}^{2,3}+\f18c_{ikjl}^{3,2}-\f78c_{ikjl}^{4,1}-\f{25}{16}c_{ikjl}^{5,0}=0
\\
\bar n_{ijkl}^{2,3}&=c_{ijkl}^{2,3}+\f32c_{ijkl}^{3,2}+\f32c_{ijkl}^{4,1}+\f5{4}c_{ijkl}^{5,0}-c_{ikjl}^{3,2}-2c_{ikjl}^{4,1}-\f52c_{ikjl}^{5,0}=0
\\
\bar n_{ijkl}^{1,5}&=c_{ijkl}^{2,2}+\f34c_{ijkl}^{3,3}+\f12c_{ijkl}^{4,2}+\f58c_{ijkl}^{5,1}+\f{15}{16}c_{ijkl}^{6,0}-\f12c_{ikjl}^{2,4}-\f14c_{ikjl}^{4,2}-\f54c_{ikjl}^{5,1}-\f{75}{32}c_{ikjl}^{6,0}=0
\\
\bar n_{ijkl}^{2,4}&=c_{ijkl}^{2,4}+\f32c_{ijkl}^{3,3}+\f32c_{ijkl}^{4,2}+\f5{4}c_{ijkl}^{5,1}+\f{15}{16}c_{ijkl}^{6,0}-c_{ikjl}^{4,2}-\f52c_{ikjl}^{5,1}-\f{15}4c_{ikjl}^{6,0}=0
\\
\bar n_{ijkl}^{3,3}&=c_{ijkl}^{3,3}+2c_{ijkl}^{4,2}+\f5{2}(c_{ijkl}^{5,1}+c_{ijkl}^{6,0})-c_{ikjl}^{3,3}-2c_{ikjl}^{4,2}-\f52(c_{ikjl}^{5,1}+c_{ikjl}^{6,0})=0    .
\label{nConstraintsEndfull}
\eal
They are mostly the same as $n^{p,q}_{ijkl}=0$ except for $(p,q)=(1,3), (1,4), (1,5),...\,$. The corresponding explicit expressions of $E^{\pm}_{p,q}$ and $F^{\pm}_{p,q}$ for the first few $\bar n^{p,q}_{ijkl}$ null constraints in 4D (remove the parts in the square brackets to get $E^{\pm}_{p,q}$ and $F^{\pm}_{p,q}$ for the $n^{p,q}_{ijkl}$ null constraints) are given by
\bal
E^+_{1,3}&=\f12 (\ell^4+2 \ell^3-7 \ell^2-8 \ell+8)+\bigg[2\bigg]\\
F^+_{1,3}&=-4 +\bigg[\f12 (-\ell^4-2 \ell^3+7 \ell^2+8 \ell-6)+1\bigg]\\
E^-_{1,3}&=\f32 (\ell^2+\ell-2)+\bigg[\f32 (\ell^2+\ell-2)\bigg]\\
F^-_{1,3}&=-2 (\ell^2+\ell-2)+\bigg[-\ell^2-\ell+2\bigg]\\
E^+_{2,2}&=\f{\ell^4}2+\ell^3-\f{7 \ell^2}2-4 \ell+6\\
F^+_{2,2}&=-\f{\ell^4}2-\ell^3+\f{7 \ell^2}2+4 \ell-6\\
E^-_{2,2}&=3 (\ell^2+\ell-2)\\
F^-_{2,2}&=-3 (\ell^2+\ell-2) \\
E^+_{1,4}&=\f1{36} (2 \ell^6+6 \ell^5-37 \ell^4-84 \ell^3+179 \ell^2+222 \ell-180)+\bigg[\f12 (-2 \ell^2-2 \ell+5)\bigg]\\
F^+_{1,4}&=-2 \ell^2-2 \ell+5+\bigg[\f1{48} (2 \ell^6+6 \ell^5-37 \ell^4-84 \ell^3+155 \ell^2+198 \ell-120)\bigg]\\
E^-_{1,4}&= \f18 (3 \ell^4+6 \ell^3-27 \ell^2-30 \ell+35)+\bigg[\f{15}8\bigg]\\
F^-_{1,4}&=-5+\bigg[\f1{16} (\ell^4+2 \ell^3-9 \ell^2-10 \ell+40)\bigg]\\
E^+_{2,3}&=\f1{36} (2 \ell^6+6 \ell^5-37 \ell^4-84 \ell^3+251 \ell^2+294 \ell-360)\\
F^+_{2,3}&=-4 \ell^2-4 \ell+10\\
E^-_{2,3}&=\f14 (3 \ell^4+6 \ell^3-27 \ell^2-30 \ell+40)\\
F^-_{2,3}&=-\f12\ell^4-\ell^3+\f92 \ell^2+5 \ell-10\\
E^+_{1,5}&=\f1{288} (\ell^8+4 \ell^7-38 \ell^6-128 \ell^5+457 \ell^4+1132 \ell^3-1860 \ell^2-2448 \ell+1728)+\bigg[\f32\bigg]\\
F^+_{1,5}&=-6+\bigg[\f1{576}(-\ell^8-4 \ell^7+38 \ell^6+128 \ell^5-457 \ell^4-1132 \ell^3+1860 \ell^2+2448 \ell-864)\bigg]\\
E^-_{1,5}&=\f1{24} (\ell^6+3 \ell^5-23 \ell^4-51 \ell^3+127 \ell^2+153 \ell-135)+\bigg[\f58 (\ell^2+\ell-3)\bigg]\\
F^-_{1,5}&=-2 (\ell^2+\ell-3)+\bigg[\f12 (-\ell^2-\ell+3)\bigg]\\
E^+_{2,4}&= \f1{288} (\ell^8+4 \ell^7-38 \ell^6-128 \ell^5+601 \ell^4+1420 \ell^3-3444 \ell^2-4176 \ell+4320)\\
F^+_{2,4}&=-\f{\ell^4}2-\ell^3+\f{11 \ell^2}2+6 \ell-15\\
E^-_{2,4}&=\f1{12} (\ell^6+3 \ell^5-23 \ell^4-51 \ell^3+142 \ell^2+168 \ell-180)\\
F^-_{2,4}&=-5 (\ell^2+\ell-3)\\
E^+_{3,3}&=\ell^4+2 \ell^3-11 \ell^2-12 \ell+20\\
F^+_{3,3}&=-\ell^4-2 \ell^3+11 \ell^2+12 \ell-20\\
E^-_{3,3}&=\f1{18} (\ell^6+3 \ell^5-23 \ell^4-51 \ell^3+202 \ell^2+228 \ell-360)\\
F^-_{3,3}&=\f1{18} (-\ell^6-3 \ell^5+23 \ell^4+51 \ell^3-202 \ell^2-228 \ell+360)    .
\eal
Note that we have artificially separated the 13, 14, 15 components of the $E$ and $F$ quantities into to two parts. The parts outside the square bracket is the $E$ and $F$ quantities corresponding to the $n^{p,q}_{ijkl}$ null constraints, that is, the parts in the square bracket is projected out if we symmetrize the $jl$ indices for $\bar n^{p,q}_{ijkl}=0$. Without the square bracket parts, the null constraints directly reduce to the null constraints for the case of identical scalar scattering in \cite{Tolley:2020gtv};  otherwise they may reduce to some superpositions of these null constraints.

\section{The case for massive fields}

\label{app:massive}

In the main text, we assumed that the hierarchy between the masses of the IR and UV particles is so large that we can treat the IR particles as effectively massless. However, this may not  aways be the case. Take one of the earliest EFT, SU(2) chiral perturbation theory for strong interactions, for example. The ratio between the pion threshold and the cutoff is only about 0.24, in which case the masses of the IR modes can introduce significant corrections to the positivity bounds \cite{Manohar:2008tc, Wang:2020jxr}. In general, if the particles are of different masses, forming the positivity bounds as a polynomial matrix program with only continuous variable appears to be elusive. This is because, for the general case, $s+t+u=m_i^2+m_j^2+m_k^2+m_l^2$ will appear in the denominator of the crossed $u$ channel, and the kinematics will now be very complicated, which leads to a very awkward expression for $\cos\thi$ in the partial wave expansion, so that the partial wave expansion depends on the particle masses $m_i,m_j,m_k,m_l$ highly nonlinearly. Nevertheless, if all the modes are of the same mass $m$ such as those in a symmetry multiplet, it is straightforward to generalize our formalism to include the mass corrections. 

We can essentially follow the same steps as the massless case. The differences are that now we define the $v$ variable and the subtraction point $\mu_p$ as
\be
v= s+\f{t}{2}-2m^2,~~~~~ \mu_p= 2m^2-\f{t}{2}    ,
\ee
which leads to the following dispersion relation for the pole-removed amplitude 
\bal
 B_{ijkl}(s,t) &= A_{ijkl}(s,t)- \frac{\li_{ijkl}}{m^2-s}  - \frac{\li_{ijkl}}{m^2 -t} - \frac{\li_{ijkl}}{m^2 -u}
\\
& = \tilde{a}^{(0)}_{ijkl}(t)+a^{(1)}_{ijkl}(t)v+ v^2 \int^\infty_{4m^2}  \frac{\d \mu}{\pi (\mu-2m^2+\f{t}2)^2} \[ \frac{{\rm Abs}\, A_{ijkl}(\mu,t)}{\mu-2m^2+\f{t}2-v}+\frac{{\rm Abs}\, A_{ilkj}(\mu,t)}{\mu-2m^2+\f{t}2+v}   \]  
\nn
& =   \tilde{a}^{(0)}_{ijkl}(t)+a^{(1)}_{ijkl}(t)v+ v^2 \int^\infty_{2m^2}  \frac{\d \mu}{\pi (\mu+\f{t}2)^2} \[ \frac{{\rm Abs}\, A_{ijkl}(\mu+2m^2,t)}{\mu+\f{t}2-v}+\frac{{\rm Abs}\, A_{ilkj}(\mu+2m^2,t)}{\mu+\f{t}2+v}   \]      ,
\nonumber
\eal
where in the last step we have shifted the integration variable by $2m^2$. Note that if desirable we could also subtract out the known low energy contribution of the dispersive integral from $4m^2$ to $\Lambda^2$ as we did in the main text. The absorptive part of the amplitude in the massive case can be expanded by partial waves as follows
\be
{\rm Abs} \, A_{ijkl}(\mu,t)=  \frac{2^{4\ai+2}\pi^\ai \Gamma(\ai)\mu^{\f12}}{{(\mu-4m^2)}^{\ai}}  \sum_{\ell=0}^\infty (2\ell+2\ai)C^{(\ai)}_{\ell}\(1+\f{2t}{\mu-4m^2}\) {\rm Abs}\, a^{ijkl}_\ell(\mu) , ~~~\ai = \frac{D-3}2   ,
\ee
Expanding both sides of the dispersion relation, we can get the sum rules for the massive case:
\be
c_{ijkl}^{m,n} = \< C_{ijkl}^{m,n}\> \equiv \<   \[m_\ell^{ij} (m_\ell^{kl})^* +(-1)^m  m_\ell^{il} (m_\ell^{kj})^*     \] \tilde C_\ell^{m,n}(\mu)  \>,~~m\geq 2,n\geq 0   ,
\ee
where $\tilde C_\ell^{m,n}$ is now given by
\be
\tilde C_\ell^{m,n}(\mu)=\sum_{p=0}^n \f{L^p_{\ell}H^{n-p}_{m+1}}{\mu^{m+1+p}(\mu-2m^2)^{n-p}}     .
\ee
With these established, we can follow the same steps as the massless case to obtain the positivity bounds with {\nb SDPB}.

\bibliographystyle{JHEP}
\bibliography{refs}

\end{document}